\documentclass{desyproc}

\def\beq{\begin{equation}}
\def\eeq{\end{equation}}
\def\beqa{\begin{eqnarray}}
\def\eeqa{\end{eqnarray}}

\begin{document}
%------------------------------------
\title{Top Quark Production}

%for single authors the superscripts are optional
\author{{\slshape Nikolaos Kidonakis}\\[1ex]
Kennesaw State University, Physics \#1202, 1000 Chastain Rd., Kennesaw, 
GA 30144, USA}

% please enter the contribution ID for the DOI
\contribID{xy}

% TO THE CONFERENCE EDITORS:
% please update the following information
% before sending the template to the authors
%\confID{999}  % if the conference is on Indico uncomment this line
\desyproc{DESY-PROC-2013-03}
\acronym{HQ2013} % if you want the Acronym in the page footer uncomment this line
%\doi  % if there is an online version we will register DOIs

\maketitle

\begin{abstract}
I discuss top quark production in hadronic collisions. I present the soft-gluon
resummation formalism and its derivation from factorization and 
renormalization-group evolution, and two-loop calculations of soft anomalous 
dimensions in the eikonal approximation. I discuss the contributions of 
next-to-next-to-leading order (NNLO) soft-gluon corrections to the total cross 
sections and top-quark transverse momentum and rapidity distributions for  
top-antitop pair production, and for single-top production in the $t$ and $s$ 
channels and in association with a $W$ boson or a charged Higgs boson. 
\end{abstract}

\section{Introduction}

The top quark is the heaviest elementary particle known to date. It was discovered in proton-antiproton collisions at the Fermilab Tevatron in 1995 in top-antitop pair production events by the CDF and D0 Collaborations \cite{CDFttb95,D0ttb95}. The uniqueness of the top quark is not only due to its heavy mass, which makes it important for Higgs physics, but also due to the fact that it is the only quark that decays before it can hadronize. Top-antitop pair and single-top production have by now been fully established at both the Tevatron and the LHC and are in good agreement with theoretical expectations, as we will see later in detail.

In these lectures I discuss top quark production in hadron colliders, paying particular attention to higher-order corrections from soft-gluon resummation. 
I begin with a discussion of higher-order soft-gluon 
corrections, factorization, renormalization-group evolution (RGE),  
resummation, and next-to-next-to-leading order (NNLO) expansions.

I continue with one- and two-loop eikonal diagrams, calculations of the 
massive cusp anomalous dimension, and presentation of the the two-loop soft 
anomalous dimension matrices for top-pair production.

I then provide results for  $t {\bar t}$ production, including the total  
$t{\bar t}$ cross sections at the LHC and the Tevatron, 
the top-quark transverse momentum, $p_ T$, distributions, and the top-quark 
rapidity distributions. Finally, I discuss   
single-top production, in particular $t$-channel and $s$-channel production,  
and $tW^-$ and $tH^-$ production, and present total cross sections and 
top-quark $p_T$ distributions. 

\section{Higher-order soft-gluon corrections}

QCD corrections are significant for hard-scattering cross-sections, 
and in particular for top-pair and single-top production.
The complete next-to-leading order (NLO) corrections were calculated 
for $t{\bar t}$ production in \cite{NLOtt1,NLOtt2} and for single-top 
production in \cite{BWHL}.

Soft-gluon corrections, i.e. perturbative corrections from the emission of soft (low-energy) gluons, originate from incomplete cancellations of infrared divergences between virtual diagrams and real diagrams with soft gluons.

The soft-gluon terms are of the form 
$\left[\frac{\ln^k(s_4/m_t^2)}{s_4}\right]_+$ 
where $k \le 2n-1$ for the $n$th-order perturbative corrections,  
and $s_4$ is the kinematical distance from partonic threshold.
The leading logarithms (LL) are those with the highest power, $2n-1$; 
the next-to-leading logarithms (NLL) have a power of one less; 
the next-to-next-to-leading logarithms (NNLL) have a power of two less, etc.
The importance of soft-gluon corrections is because they are dominant near 
threshold. It is possible to resum (i.e. exponentiate) these corrections to all orders in perturbative QCD. This resummation follows from factorization of the cross section and RGE of its factors. The resummation of the leading logarithms requires universal terms describing the emission of collinear and soft gluons that only depend on the identity of the incoming and outgoing partons, i.e. the details of the hard process are irrelevant. However, at NLL accuracy \cite{NKGS} and beyond it is necessary to involve the process-dependent color exchange in the hard-scattering process and to perform the corresponding loop calculations in the eikonal approximation.

In addition to these soft-gluon logarithmic terms there also arise 
terms of purely collinear origin, of the form $\frac{1}{m_t^2} \ln^k(s_4/m_t^2)$, but we will not dicuss these kind of terms in this paper.

Complete results are now available at NNLL accuracy, which requires the calculation of two-loop soft anomalous dimensions. For a review of resummation for top quark production see Ref. \cite{NKBP}.
Approximate next-to-next-to-leading order (NNLO) double-differential cross 
sections and even next-to-next-to-next-to-leading order (NNNLO) corrections 
have been derived from the expansion of the resummed results \cite{NKNNNLO}.

\subsection{Factorization, RGE, and Resummation}

We consider hadronic processes of the form
\beqa
h_1(p_{h_1})+h_2(p_{h_2}) \rightarrow t(p)+X 
\nonumber
\eeqa
where $h_1$, $h_2$, are colliding hadrons (protons at the LHC; protons and antiprotons at the Tevatron) and $t$ denotes the observed top quark with $X$ all additional final-state particles. 
The underlying partonic processes are of the form
\beqa
f_{1}(p_1)\, + \, f_{2}\, (p_2) \rightarrow t(p)\, + \, X 
\nonumber
\eeqa
where $f_1$ and $f_2$ represent partons (quarks or gluons).
We define $s=(p_1+p_2)^2$, $t=(p_1-p)^2$, $u=(p_2-p)^2$. 
Also $s_4=s+t+u-\sum m^2$, where the sum is over the squared masses 
of all particles in the process. Thus, $s_4$ measures distance from partonic 
threshold, where there is no energy for additional radiation, but the top quark 
may have arbitrary momentum and is not restricted to be produced at rest (thus partonic threshold is more general than absolute, or production, threshold where the top quark is produced at rest). At partonic threshold $s_4=0$.

The factorization for the (in general differential) cross section is expressed by the formula 
\beqa
d\sigma_{h_1 h_2\rightarrow tX}&=&
\sum_{f_1,f_2} \; 
\int dx_1 \, dx_2 \,  \phi_{f_1/h_1}(x_1,\mu_F) \, 
\phi_{f_2/h_2}(x_2,\mu_F)\, 
{\hat \sigma}_{f_1 f_2 \rightarrow tX}(s_4,s,t,u,\mu_F,\mu_R)
\nonumber
\eeqa
where $\phi$ are parton distribution functions with $x_1$ and $x_2$ the momentum fractions of partons $f_1$ and $f_2$ in hadrons $h_1$ and $h_2$ respectively, $\mu_F$ is the factorization 
scale and $\mu_R$ is the renormalization scale.
We factorize the initial-state collinear divergences into the parton 
distribution functions, $\phi$.
Soft-gluon corrections appear in the partonic hard-scattering cross section, 
${\hat \sigma}_{f_1 f_2 \rightarrow tX}$,  
as plus distributions of logarithmic terms, 
defined through their integral with parton distribution functions 
\beqa
\int_0^{s_{4 \, max}} ds_4 \, \phi(s_4) \left[\frac{\ln^k(s_4/m_t^2)}
{s_4}\right]_{+} &\equiv&
\int_0^{s_{4\, max}} ds_4 \frac{\ln^k(s_4/m_t^2)}{s_4} [\phi(s_4) - \phi(0)]
\nonumber \\ &&
{}+\frac{1}{k+1} \ln^{k+1}\left(\frac{s_{4\, max}}{m_t^2}\right) \phi(0) \, .
\nonumber
\eeqa

Resummation follows from the factorization properties of the 
cross section, performed in moment space.
We define moments of the partonic cross section by  
${\hat\sigma}(N)=\int (ds_4/s) \;  e^{-N s_4/s} {\hat\sigma}(s_4)$. 
The logarithms of $s_4$ give rise to logarithms of $N$ in moment space,  
and we will show that the logarithms of $N$ appearing in ${\hat \sigma}(N)$ 
exponentiate.

We then write a factorized expression for the infrared-regularized 
(with $\epsilon=4-n$) 
parton-parton scattering cross section,  
$\sigma_{f_1 f_2 \rightarrow tX}(N, \epsilon)$, in moment space
\beqa
\sigma_{f_1 f_2 \rightarrow tX}(N, \epsilon)  
=\phi_{f_1/f_1}(N,\mu_F,\epsilon)\; 
\phi_{f_2/f_2}(N,\mu_F,\epsilon) \;
{\hat \sigma}_{f_1 f_2 \rightarrow tX}(N,\mu_F,\mu_R) 
\nonumber
\eeqa
which factorizes similarly to the hadronic cross section, 
with  $\phi(N)=\int_0^1dx\; x^{N-1}\phi(x)$. 

\begin{figure}[hb]
\centerline{\includegraphics[width=0.40\textwidth]{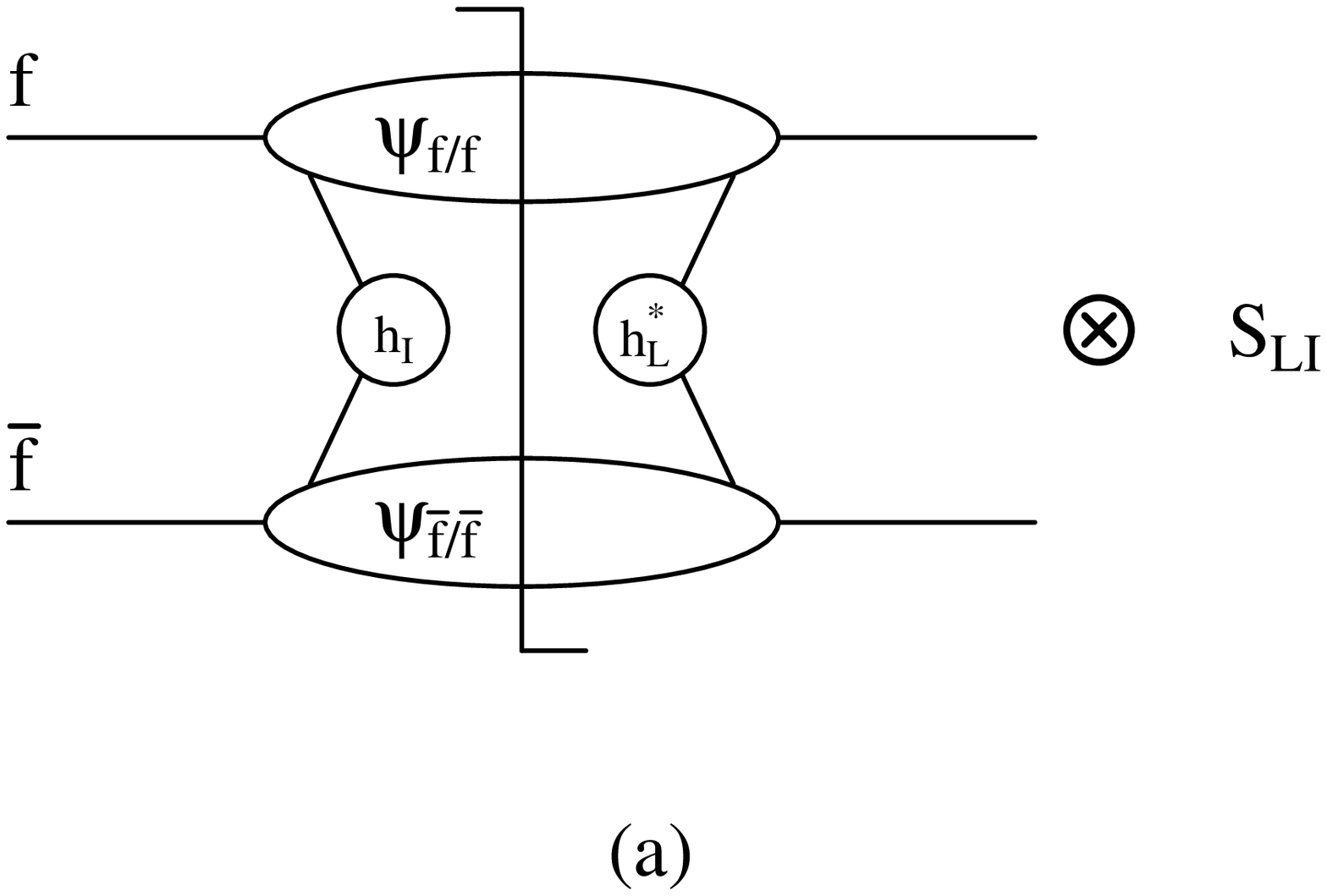}
\hspace{10mm}\includegraphics[width=0.40\textwidth]{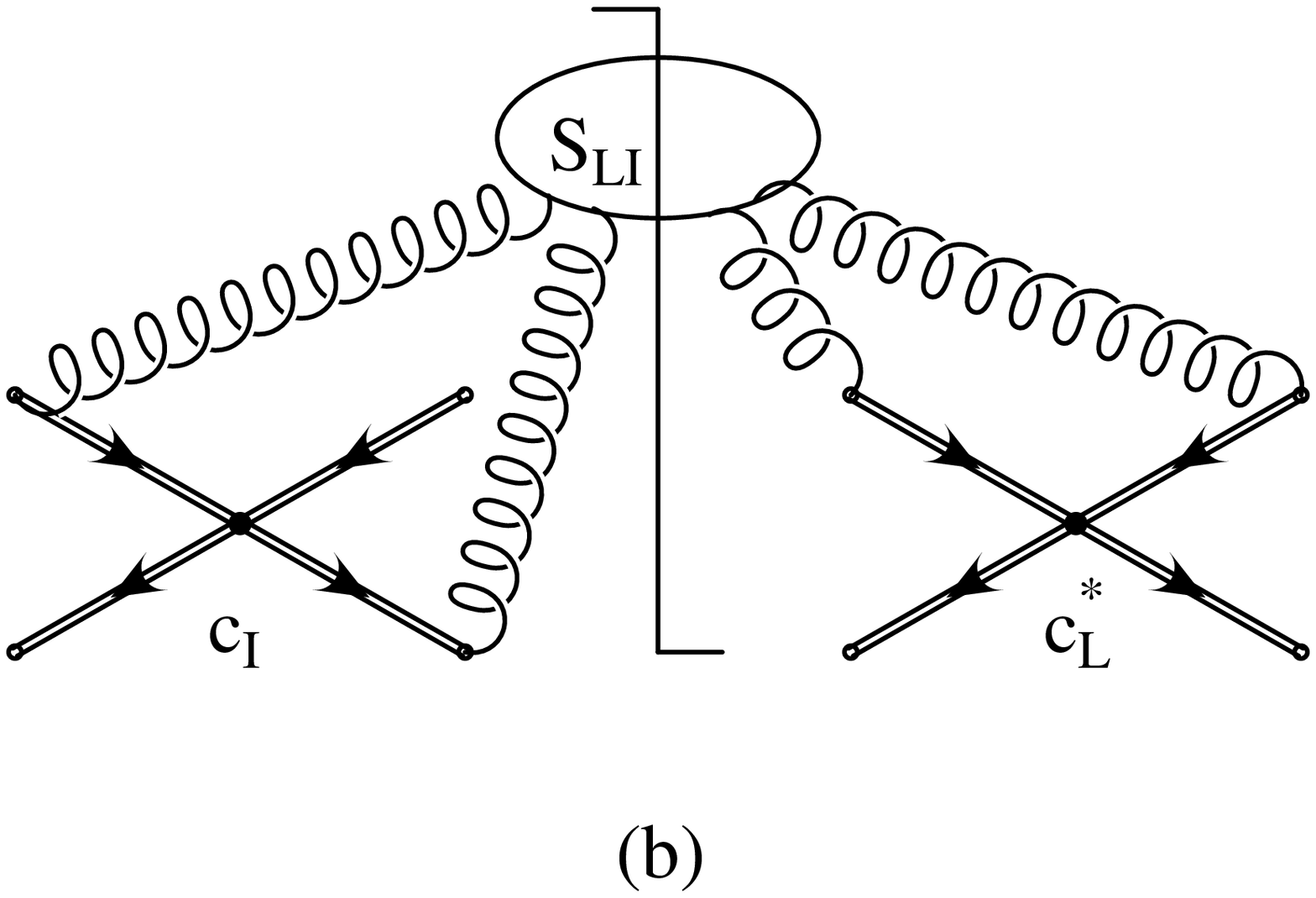}}
\caption{Factorization of the partonic cross section for $t{\bar t}$ production: (a) The functions involved in the partonic process; (b) the soft-gluon function $S$.}
\label{fig1}
\end{figure}

The partonic function ${\hat \sigma}_{f_1 f_2 \rightarrow tX}$ still has sensitivity
to soft-gluon dynamics via its $N$ dependence.   
We then refactorize the cross section \cite{NKGS} in terms of modified parton 
distributions $\psi$, defined in the partonic center-of-mass frame at fixed 
energy, as
\beqa
\sigma_{f_1 f_2\rightarrow tX}(N,\epsilon)
&=&\psi_{f_1/f_1} \left(N,\mu_F,\epsilon \right) \;
\psi_{f_2/f_2} \left(N,\mu_F,\epsilon \right)  
\nonumber \\ && \times \; 
H_{IL}^{f_1 f_2\rightarrow tX} \left(\alpha_s(\mu_R)\right)\; 
S_{LI}^{f_1 f_2 \rightarrow tX} 
\left(\frac{m_t}{N \mu_F},\alpha_s(\mu_R) \right)\;
\prod_j  J_j\left (N,\mu_F,\epsilon \right)  \, .
\nonumber
\eeqa 
This factorization is shown for the case of top-antitop pair production in Fig. \ref{fig1}(a).

The $H_{IL}^{f_1 f_2\rightarrow tX}$ are $N$-independent hard-scattering terms 
which involve contributions 
from the amplitude of the process and the complex conjugate of the amplitude,
in the form $H_{IL}=h_L^*\, h_I$. 
Also, $S_{LI}^{f_1 f_2\rightarrow tX}$ is the soft gluon function for non-collinear 
soft-gluon emission;
it represents the coupling of soft gluons to the
partons in the scattering with color tensors $c_I$, $c_L$ 
(see Fig. \ref{fig1}(b)). Both $H_{IL}$ and $S_{LI}$ are process dependent and 
they are matrices in the space of color exchanges in the partonic scattering.
$J$ are jet functions describing universal soft and collinear emission from 
any outgoing massless partons.

Comparing the two previous equations, we find
\beqa
{\hat \sigma}_{f_1 f_2\rightarrow tX}(N,\mu_F,\mu_R)&=&
\frac{\psi_{f_1/f_1}(N,\mu_F,\epsilon)\, 
\psi_{f_2/f_2}(N,\mu_F,\epsilon)}
{\phi_{f_1/f_1}(N,\mu_F,\epsilon)\, 
\phi_{f_2/f_2}(N,\mu_F,\epsilon)}
\nonumber \\ &&  \times \;
H_{IL}^{f_1 f_2\rightarrow tX}\left(\alpha_s(\mu_R)\right) \, 
S_{LI}^{f_1 f_2\rightarrow tX}
\left(\frac{m_t}{N\mu_F},\alpha_s(\mu_R)\right) 
\prod_j  J_j\left (N,\mu_F,\epsilon \right) \, .
\nonumber
\eeqa

All the factors in the above equation are gauge and factorization
scale dependent. The requirement that the product of these factors be
independent of the gauge and the factorization scale results
in the exponentiation of logarithms of $N$ in the ratios 
$\psi_{f_1/f_1}/\phi_{f_1/f_1}$ and $\psi_{f_2/f_2}/\phi_{f_2/f_2}$, 
in the soft-gluon matrix $S_{LI}$, and in the functions $J_j$. 
 
The soft matrix $S_{LI}$ requires renormalization as a composite operator;  
its $N$-dependence can then be resummed via RGE \cite{NKGS}.
The product $H_{IL}S_{LI}$ however needs
no overall renormalization, because the UV divergences of $H_{IL}$
balance those of $S_{LI}$. We have 
\beqa
H^b_{IL}&=& \left(\prod_{i=1,2} Z_i^{-1} \right)\; \left(Z_S^{-1}\right)_{IA}
H_{AB} \left[\left(Z_S^\dagger \right)^{-1}\right]_{BL} 
\nonumber \\ 
S^b_{LI}&=&(Z_S^\dagger)_{LC}S_{CD}Z_{S,DI}
\nonumber
\eeqa
where $H^b$ and $S^b$ are the unrenormalized quantities,
$Z_i$ are the renormalization constants of the 
incoming partonic fields, and $Z_S$ is
a matrix of renormalization constants, which describe the
renormalization of the soft function, including the
wave functions for outgoing heavy-quark eikonal lines.

Thus  $S_{LI}$ satisfies the renormalization group equation
\beqa
\left(\mu \frac{\partial}{\partial \mu}
+\beta(g_s)\frac{\partial}{\partial g_s}\right)\,S_{LI}
=-(\Gamma^\dagger_S)_{LC}S_{CI}-S_{LD}(\Gamma_S)_{DI}
\nonumber
\eeqa
where $g_s^2=4\pi\alpha_s$ and 
$\beta$ is the QCD beta function
\beqa
\beta(\alpha_s) \equiv \frac{1}{2\alpha_s}\frac{d\alpha_s}{d\ln \mu}
=\mu \, d \ln g_s/d \mu
=-\beta_0 \alpha_s/(4 \pi)-\beta_1 \alpha_s^2/(4 \pi)^2+\cdots \, ,
\nonumber
\eeqa
with $\beta_0=(11C_A-2n_f)/3$ and $\beta_1=34 C_A^2/3-2n_f(C_F+5C_A/3)$.
Here $C_F=(N_c^2-1)/(2N_c)$ and $C_A=N_c$, with $N_c=3$ the number of colors, 
and  $n_f$ is the number of light quark flavors ($n_f=5$ for top production).

$\Gamma_S$ is the soft anomalous dimension matrix 
that controls the evolution of the soft function $S$. 
In dimensional regularization $Z_S$ has $1/\epsilon$ poles, and $\Gamma_S$
is given at one loop in terms of the residue of $Z_S$ by
\beqa
\Gamma_S^{(1-loop)} (g_s)=-\frac{g_s}{2} \frac {\partial}{\partial g_s}
{\rm Res}_{\epsilon\rightarrow 0} Z_S (g_s, \epsilon) \, .
\nonumber
\eeqa

The soft anomalous dimension $\Gamma_S$ is a matrix in 
color space and a function of the kinematical invariants $s$, $t$, $u$.
The process-dependent matrices $\Gamma_S$ have been 
calculated at one loop for all $2 \rightarrow 2$ 
partonic processes.
For the $q{\bar q} \rightarrow t{\bar t}$ process,  $\Gamma_S$ is a $2\times 2$ 
matrix \cite{NKGS}.
For the  $gg \rightarrow t{\bar t}$ process, $\Gamma_S$  is a $3\times 3$ matrix \cite{NKGS}. Explicit expressions at one and two loops will be provided in 
Section 4.

The resummed cross section in moment space, denoted as  
${\hat{\sigma}}_{f_1 f_2\rightarrow tX}^{res}(N)$ below, 
follows from the RGE of all the functions in the factorized cross section, and 
can be written in the form:
\beqa
{\hat{\sigma}}_{f_1 f_2\rightarrow tX}^{res}(N) &=&
\exp\left[ \sum_{i=1,2} E_i(N_i)\right] \, \exp\left[ \sum_j E'_j(N')\right]\;
\exp \left[\sum_{i=1,2} 2 \int_{\mu_F}^{\sqrt{s}} \frac{d\mu}{\mu}\;
\gamma_{i/i}\left({\tilde N}_i, \alpha_s(\mu)\right)\right] \;
\nonumber\\ && \times \,
{\rm tr} \left\{H^{f_1 f_2\rightarrow tX}\left(\alpha_s(\sqrt{s})\right)
\exp \left[\int_{\sqrt{s}}^{{\sqrt{s}}/{\tilde N'}}
\frac{d\mu}{\mu} \;
\Gamma_S^{\dagger \, f_1 f_2\rightarrow tX}\left(\alpha_s(\mu)\right)\right] \right.
\nonumber\\ && \left. \times \,
S^{f_1 f_2\rightarrow tX} \left(\alpha_s\left(\frac{\sqrt{s}}{\tilde N'}\right)
\right) \;
\exp \left[\int_{\sqrt{s}}^{{\sqrt{s}}/{\tilde N'}}
\frac{d\mu}{\mu}\; \Gamma_S^{f_1 f_2\rightarrow tX}
\left(\alpha_s(\mu)\right)\right] \right\}
\nonumber 
\eeqa
where the trace is taken of the product of the color-space matrices $H$, $S$, 
and exponents of $\Gamma_S$ and its Hermitian conjugate, $\Gamma_S^{\dagger}$.

The collinear and soft radiation from incoming partons is resummed via the first exponential with 
\beqa
E_i(N_i)&=&
\int^1_0 dz \frac{z^{N_i-1}-1}{1-z}\;
\left \{\int_1^{(1-z)^2} \frac{d\lambda}{\lambda}
A_i\left(\alpha_s(\lambda s)\right)
+D_i\left[\alpha_s((1-z)^2 s)\right]\right\} 
\nonumber
\eeqa
(for purely collinear corrections,  replace $\frac{z^{N-1}-1}{1-z}\;$ by $\;- z^{N-1}$ in the above expression).
Here $N_1=N(m_t^2-u)/m_t^2$ and $N_2=N(m_t^2-t)/m_t^2$. The term $A_i$ has the 
perturbative expansion 
$A_i = \frac{\alpha_s}{\pi} A_i^{(1)}
+\left(\frac{\alpha_s}{\pi}\right)^2 A_i^{(2)}+\cdots$
where
$A_i^{(1)}=C_i$ \cite{GS87} with $C_i=C_F$ for a quark 
or antiquark and $C_i=C_A$ for a gluon, 
while $A_i^{(2)}=C_i K/2$ \cite{CT89} with 
$K= C_A\; ( 67/18-\zeta_2 ) - 5n_f/9$ \cite{KT82}. 
Here and below we use $\zeta_2=\pi^2/6$, $\zeta_3=1.2020569\cdots$,  
and $\zeta_4=\pi^4/90$. 
Also $D_i=(\alpha_s/\pi)D_i^{(1)}+(\alpha_s/\pi)^2 D_i^{(2)}+\cdots$
with $D_i^{(1)}=0$ in Feynman gauge ($D_i^{(1)}=-C_i$ in axial gauge).
In Feynman gauge the two-loop result is \cite{CLS97}
\beqa
D_i^{(2)}=C_i C_A \left(-\frac{101}{54}+\frac{11}{6} \zeta_2
+\frac{7}{4}\zeta_3\right)
+C_i n_f \left(\frac{7}{27}-\frac{\zeta_2}{3}\right) \, .
\nonumber
\eeqa

The collinear and soft radiation from outgoing massless quarks and gluons is resummed via the second exponential 
\beqa
E_j'(N') &=& 
\int^1_0 dz \frac{z^{N'-1}-1}{1-z}
\left \{\int^{1-z}_{(1-z)^2} \frac{d\lambda}{\lambda}
A_j \left(\alpha_s\left(\lambda s\right)\right)
+B_j \left[\alpha_s((1-z)s)\right] \right.
\nonumber \\ && \hspace{28mm} \left.
+D_j \left[\alpha_s((1-z)^2 s)\right]\right\} 
\nonumber
\eeqa
where $N'=N \, s/m_t^2$. 
Note that this exponent is not needed in $t{\bar t}$ production but it is 
used in single-top production.
The term $B_j$ has the perturbative expansion
$B_j=(\alpha_s/\pi)B_j^{(1)}+(\alpha_s/\pi)^2 B_j^{(2)}+\cdots$
with $B_q^{(1)}=-3C_F/4$ for a quark or antiquark, and $B_g^{(1)}=-\beta_0/4$ 
for a gluon \cite{GS87,CT89}.
Also (c.f. \cite{CLS97,MVV})
\beqa
B_q^{(2)}=C_F^2\left(-\frac{3}{32}+\frac{3}{4}\zeta_2-\frac{3}{2}\zeta_3\right)
+C_F C_A \left(-\frac{57}{32}-\frac{11}{12}\zeta_2+\frac{3}{4}\zeta_3\right)
+n_f C_F \left(\frac{5}{16}+\frac{\zeta_2}{6}\right) \, ,
\nonumber
\eeqa
\beqa
B_g^{(2)}=C_A^2\left(-\frac{1025}{432}-\frac{3}{4}\zeta_3\right)
+\frac{79}{108} C_A \, n_f +C_F \frac{n_f}{8}-\frac{5}{108} n_f^2 \, .
\nonumber
\eeqa

The factorization scale dependence in the third exponential is controlled by
the moment-space anomalous dimension of the ${\overline {\rm MS}}$ density 
$\phi_{i/i}$, which is $\gamma_{i/i}=-A_i \ln {\tilde N}_i +\gamma_i$ 
\cite{GALY79,GFP80}, where ${\tilde N}_i=N_i e^{\gamma_E}$ with $\gamma_E$ the 
Euler constant.
The parton anomalous dimensions $\gamma_i$  
have the perturbative expansion 
\beqa
\gamma_i=(\alpha_s/\pi) \gamma_i^{(1)}
+(\alpha_s/\pi)^2 \gamma_i^{(2)} + \cdots
\nonumber
\eeqa
with  $\gamma_q^{(1)}=3C_F/4$, $\gamma_g^{(1)}=\beta_0/4$,
\beqa
\gamma_q^{(2)}=C_F^2\left(\frac{3}{32}-\frac{3}{4}\zeta_2
+\frac{3}{2}\zeta_3\right)
+C_F C_A\left(\frac{17}{96}+\frac{11}{12}\zeta_2-\frac{3}{4}\zeta_3\right)
+n_f C_F \left(-\frac{1}{48}-\frac{\zeta_2}{6}\right)
\nonumber
\eeqa
and
\beqa
\gamma_g^{(2)}=C_A^2\left(\frac{2}{3}+\frac{3}{4}\zeta_3\right)
-n_f\left(\frac{C_F}{8}+\frac{C_A}{6}\right) \, .
\nonumber
\eeqa

The relation between $\alpha_s$ at two different scales, $\mu$ and 
$\mu_R$, is
\beqa
\alpha_s(\mu)&=&\alpha_s(\mu_R)\left[1-\frac{\beta_0}{4\pi}\alpha_s(\mu_R)
\ln\left(\frac{{\mu}^2}{\mu_R^2}\right)
+\frac{\beta_0^2}{16\pi^2}\alpha_s^2(\mu_R)
\ln^2\left(\frac{{\mu}^2}{\mu_R^2}\right)
-\frac{\beta_1}{16\pi^2}\alpha_s^2(\mu_R)
\ln\left(\frac{{\mu}^2}{\mu_R^2}\right) \right.
\nonumber \\ && \hspace{20mm} \left.
+\cdots\right] \, .
\nonumber
\eeqa

We write the perturbative expansions for the hard-scattering function $H$ and 
the soft-gluon function $S$ as  
\beqa
H=\alpha_s^{d_{\alpha_s}}H^{(0)}+\frac{\alpha_s^{d_{\alpha_s}+1}}{\pi} H^{(1)}
+\frac{\alpha_s^{d_{\alpha_s}+2}}{\pi^2} H^{(2)}+\cdots
\nonumber
\eeqa
and
\beqa
S=S^{(0)}+\frac{\alpha_s}{\pi} S^{(1)}+\frac{\alpha_s^2}{\pi^2} S^{(2)}+\cdots
\nonumber
\eeqa 
respectively, where $d_{\alpha_s}$ denotes the power of $\alpha_s$ in the Born 
cross section.
At lowest order, the trace of the product of the hard matrices
$H$ and soft matrices $S$ gives the Born cross section for each partonic 
process, $\sigma^B=\alpha_s^{d_{\alpha_s}}{\rm tr}[H^{(0)}S^{(0)}]$.

Noncollinear soft gluon emission is controlled by the soft anomalous dimension 
$\Gamma_S$, which has the perturbative expansion 
\beqa
\Gamma_S=\frac{\alpha_s}{\pi} \Gamma_S^{(1)}
+\frac{\alpha_s^2}{\pi^2} \Gamma_S^{(2)}+\cdots
\nonumber
\eeqa

We determine $\Gamma_S$ from the coefficients of ultraviolet poles 
in dimensionally regularized eikonal diagrams.
The determination of $\Gamma_S^{(1)}$ 
is needed for NLL resummation and it requires one-loop calculations in 
the eikonal approximation; $\Gamma_S^{(2)}$ is needed for NNLL resummation and 
requires two-loop calculations. 

Complete two-loop results are now known for the soft anomalous dimensions for many processes, and in these lectures I will review results for:

$\bullet$ the soft (cusp) anomalous dimension for 
$e^+ e^- \rightarrow t {\bar t}$

$\bullet$ $t{\bar t}$ hadroproduction

$\bullet$ $t$-channel single top production 

$\bullet$ $s$-channel single top production 

$\bullet$ $bg \rightarrow t W^-$ and $bg \rightarrow t H^-$

\subsection{NLO and NNLO expansions}

The resummed cross section suffers from infrared divergences that need a 
prescription to be dealt with. However, the numerical results depend on the prescription, and differences between prescriptions are typically larger than corrections beyond NNLO. Thus, an alternative and preferred procedure is to 
expand the resummed cross section to a fixed order in the perturbative expansion, thus avoiding arbitrary prescription dependences. Thus the resummed cross section is used as a generator of higher-order soft-gluon corrections, and here we present expansions to NNLO (for NNNLO see the second paper in \cite{NKNNNLO}).

In the moment-space resummed cross section we are resumming $\ln^k N$; we then expand to fixed order and invert back to momentum space to get the usual $\ln^k(s_4/m_t^2)/s_4$ terms.

We will use the following notation for the logarithmic plus distributions,  
\beqa
{\cal D}_k(s_4)\equiv\left[\frac{\ln^k(s_4/m_t^2)}{s_4}\right]_+ \, .
\nonumber
\eeqa

The NLO soft-gluon corrections from the expansion of the resummed cross section can be written as 
\beqa
{\hat{\sigma}}^{(1)} = \sigma^B \frac{\alpha_s(\mu_R)}{\pi}
\left\{c_3\, {\cal D}_1(s_4) + c_2\,  {\cal D}_0(s_4) 
+c_1\,  \delta(s_4)\right\}+\frac{\alpha_s^{d_{\alpha_s}+1}(\mu_R)}{\pi} 
\left[A^c \, {\cal D}_0(s_4)+T_1^c \, \delta(s_4)\right] 
\nonumber
\eeqa
where we have separated contributions into a part proportional to the Born term, i.e. the leading-order (LO) term, $\sigma^B$, and a part that is not (in general) proportional to it. The leading logarithmic coefficient is 
\beqa
c_3=\sum_i 2 \, A_i^{(1)} -\sum_j A_j^{(1)} \, ,
\nonumber
\eeqa
and is always multiplied by $\sigma^B$. The next-to-leading logarithmic 
terms are in general not all proportional to $\sigma^B$ and are separated into 
two parts.    
The first part has coefficient $c_2$ which is defined by $c_2=c_2^{\mu}+T_2$, 
with
$c_2^{\mu}=-\sum_i A_i^{(1)} \ln(\mu_F^2/m_t^2)$
denoting the terms involving logarithms of the scale, and  
\beqa
T_2=\sum_i \left[
-2 \, A_i^{(1)} \, \ln\left(\frac{-t_i}{m_t^2}\right)+D_i^{(1)}
-A_i^{(1)} \ln\left(\frac{m_t^2}{s}\right)\right]
+\sum_j \left[B_j^{(1)}+D_j^{(1)}
-A_j^{(1)} \, \ln\left(\frac{m_t^2}{s}\right)\right] \, . 
\nonumber
\eeqa
The part not in general proportional to $\sigma^B$ is defined by 
\beqa
A^c={\rm tr} \left(H^{(0)} \Gamma_S^{(1)\,\dagger} S^{(0)}
+H^{(0)} S^{(0)} \Gamma_S^{(1)}\right) \, .
\nonumber
\eeqa

The terms proportional to $\delta(s_4)$ include virtual corrections which 
cannot be determined from resummation as well as some terms that involve 
logarithms of the scales $\mu_F$ and $\mu_R$ and which can be calculated 
from the expansion of the resummed cross section. We write
$c_1 =c_1^{\mu} +T_1$  with $c_1^{\mu}$ denoting the terms involving 
logarithms of the scale
\beqa
c_1^{\mu}=\sum_i \left[A_i^{(1)}\, \ln\left(\frac{-t_i}{m_t^2}\right) 
-\gamma_i^{(1)}\right]\ln\left(\frac{\mu_F^2}{m_t^2}\right)
+d_{\alpha_s} \frac{\beta_0}{4} \ln\left(\frac{\mu_R^2}{m_t^2}\right) \, .
\nonumber
\eeqa
However $T_1$ as well as $T_1^c$ can only be found from a complete NLO calculation.

The NNLO soft-gluon corrections from the expansion of the resummed cross section are then 
\beqa
{\hat{\sigma}}^{(2)}&=&\sigma^B \frac{\alpha_s^2(\mu_R)}{\pi^2}
\left\{\frac{1}{2}c_3^2\, {\cal D}_3(s_4) + 
\left[\frac{3}{2}c_3 c_2-\frac{\beta_0}{4} c_3
+\sum_j \frac{\beta_0}{8} A_j^{(1)}\right]  {\cal D}_2(s_4) \right.
\nonumber \\ && \hspace{-10mm}  
{}+\left[c_3 c_1+c_2^2-\zeta_2 c_3^2
-\frac{\beta_0}{2} T_2+\frac{\beta_0}{4} c_3 
\ln\left(\frac{\mu_R^2}{m_t^2}\right)+\sum_i 2A_i^{(2)}-\sum_j A_j^{(2)}
+\sum_j \frac{\beta_0}{4} B_j^{(1)}\right] {\cal D}_1(s_4)
\nonumber \\ &&  \hspace{-10mm}
{}+\left[c_2 c_1-\zeta_2 c_3 c_2+\zeta_3 c_3^2
+\frac{\beta_0}{4} c_2 \ln\left(\frac{\mu_R^2}{s}\right) \right. 
-\sum_i \frac{\beta_0}{2} A_i^{(1)} \ln^2\left(\frac{-t_i}{m_t^2}\right)
\nonumber \\ && \hspace{-5mm} 
{}+\sum_i [\left(-2 A_i^{(2)}+\frac{\beta_0}{2} D_i^{(1)}\right) 
\ln\left(\frac{-t_i}{m_t^2}\right)+D_i^{(2)}
+\frac{\beta_0}{8}  A_i^{(1)} \ln^2\left(\frac{\mu_F^2}{s}\right) 
-A_i^{(2)} \ln\left(\frac{\mu_F^2}{s}\right)] 
\nonumber \\ && \hspace{-5mm} 
{}+\sum_j [B_j^{(2)}+D_j^{(2)}-\left(A_j^{(2)}+\frac{\beta_0}{4}
(B_j^{(1)}+2 D_j^{(1)})\right) \ln\left(\frac{m_t^2}{s}\right) 
\nonumber \\ && \hspace{5mm} \left. \left.  
{}+\frac{3\beta_0}{8} A_j^{(1)} \ln^2\left(\frac{m_t^2}{s}\right)]\right] 
 {\cal D}_0(s_4) \right\}
\nonumber \\ && \hspace{-10mm}
{}+\frac{\alpha_s^{d_{\alpha_s}+2}(\mu_R)}{\pi^2} 
\left\{\frac{3}{2} c_3 A^c \, {\cal D}_2(s_4)
+\left[\left(2 c_2-\frac{\beta_0}{2}\right) A^c+c_3 T_1^c +F^c\right]
{\cal D}_1(s_4) \right. 
\nonumber \\ && \hspace{-5mm} \left. 
{}+\left[\left(c_1-\zeta_2 c_3+\frac{\beta_0}{4}\ln\left(\frac{\mu_R^2}{s}
\right)\right)A^c+c_2 T_1^c +F^c \ln\left(\frac{m_t^2}{s}\right) +G^c\right]
{\cal D}_0(s_4) \right\} 
\nonumber
\eeqa
where
$$
F^c={\rm tr} \left[H^{(0)} \left(\Gamma_S^{(1)\,\dagger}\right)^2 S^{(0)}
+H^{(0)} S^{(0)} \left(\Gamma_S^{(1)}\right)^2
+2 H^{(0)} \Gamma_S^{(1)\,\dagger} S^{(0)} \Gamma_S^{(1)} \right] $$
\beqa
G^c&=&{\rm tr} \left[H^{(1)} \Gamma_S^{(1)\,\dagger} S^{(0)}
+H^{(1)} S^{(0)} \Gamma_S^{(1)} + H^{(0)} \Gamma_S^{(1)\,\dagger} S^{(1)}
+H^{(0)} S^{(1)} \Gamma_S^{(1)} \right.
\nonumber \\ && \quad \quad \left.
{}+H^{(0)} \Gamma_S^{(2)\,\dagger} S^{(0)}
+H^{(0)} S^{(0)} \Gamma_S^{(2)} \right] 
\nonumber
\eeqa
and $c_3$, $c_2$, $c_1$, etc are from the NLO expansion. 
The two-loop universal quantities $A^{(2)}$, $B^{(2)}$, $D^{(2)}$ were given previously.
The two-loop process-dependent $\Gamma_S^{(2)}$ have been recently calculated for several processes, including top quark production in various channels. 

In addition to the plus distributions, the factorization and renormalization scale dependent terms proportional to $\delta(s_4)$ at NNLO have also been calculated \cite{NKNNNLO}. 

\section{Two-loop calculations for the massive cusp anomalous dimension}

\begin{figure}
\centerline{\includegraphics[width=0.5\textwidth]{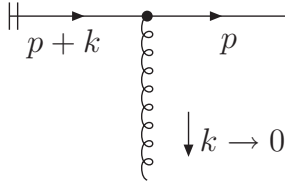}}
\caption{Elementary eikonal diagram for soft-gluon emission from an outgoing quark.}
\label{eikdiag}
\end{figure}

The Feynman rules for diagrams with soft gluon emission, 
see Fig. \ref{eikdiag}, simplify as
\beqa
{\bar u}(p) \, (-i g_s T_F^c) \, \gamma^{\mu}
\frac{i (p\!\!/+k\!\!/+m)}{(p+k)^2
-m^2+i\epsilon} \rightarrow {\bar u}(p)\,  g_s T_F^c \, \gamma^{\mu}
\frac{p\!\!/+m}{2p\cdot k+i\epsilon}
&=&{\bar u}(p)\, g_s T_F^c \,
\frac{v^{\mu}}{v\cdot k+i\epsilon}
\nonumber 
\eeqa
with ${\bar u}$ a Dirac spinor, $T_F^c$ the generators of SU(3), and $p \propto v$, for example 
we may take $p^{\mu}=\sqrt{\frac{s}{2}} v^{\mu}$, though other choices are possible.
This is the eikonal approximation.

We perform the calculations here in momentum space and Feynman gauge.
The first soft anomalous dimension that we consider is the massive cusp 
anomalous dimension, which is also the soft anomalous dimension for the 
process $e^+ e^- \rightarrow t{\bar t}$ \cite{NKPRL,NK2lproc}.

\subsection{One-loop cusp anomalous dimension}

\begin{figure}
\centerline{\includegraphics[width=0.5\textwidth]{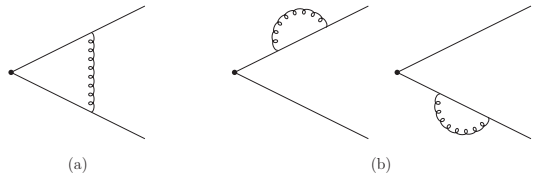}}
\caption{One-loop eikonal diagrams for the cusp anomalous dimension.}
\label{loopg1}
\end{figure}

The one-loop eikonal diagrams for the cusp anomalous dimension are shown in 
Fig. \ref{loopg1}. The eikonal lines represent the top and the antitop quarks. 
The one-loop vertex correction is graph (a) and the 
one-loop top and antitop self-energy diagrams are the two graphs (b).

The one-loop soft anomalous dimension, $\Gamma_S^{(1)}$, can be read 
off the coefficient of the ultraviolet (UV) pole of the one-loop diagrams. 
The calculation gives \cite{NKPRL,NK2lproc} 
\beqa
\Gamma_S^{(1)}=C_F \left[-\frac{(1+\beta^2)}{2\beta} 
\ln\left(\frac{1-\beta}{1+\beta}\right) -1\right]
\nonumber
\eeqa
with $\beta=\sqrt{1-\frac{4m_t^2}{s}}$.

\begin{figure}[hb]
\centerline{\includegraphics[width=0.25\textwidth]{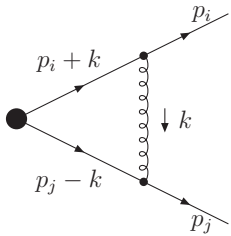}}
\caption{One-loop vertex-correction diagram.}
\label{loopv1}
\end{figure}

As an example of the calculation we provide some details for the vertex 
correction graph, i.e. diagram (a) of Fig. \ref{loopg1}. This one-loop vertex 
correction is shown in more detail, and with momenta assignments, in 
Fig. \ref{loopv1}. The integral corresponding to this diagram is 
\beqa
I_{1a} = g_s^2 \int\frac{d^n k}{(2\pi)^n} \frac{(-i)g^{\mu \nu}}{k^2} 
\frac{v_i^{\mu}}{v_i\cdot k} \, \frac{(-v_j^{\nu})}{(-v_j\cdot k)} 
\nonumber 
\eeqa
which has three factors. The first factor is the gluon propagator and the 
last two are the eikonal rules for the two lines.
Using Feynman parameterization, this can be rewritten as
\beqa
I_{1a} =-2i g_s^2 \, \frac{v_i \cdot v_j}{(2\pi)^n}
\int_0^1 dx \int_0^{1-x} dy \int\frac{d^n k} 
{\left[x k^2+y v_i \cdot k+(1-x-y) v_j \cdot k \right]^3}
\nonumber
\eeqa
which, after the integration over $k$, gives 
\beqa
I_{1a} &=& g_s^2 \, v_i \cdot v_j \, 2^{6-2n} \, \pi^{-n/2} \, 
\Gamma\left(3-\frac{n}{2}\right) \, 
\int_0^1 dx \, x^{3-n} 
\nonumber \\ && \times  
\int_0^{1-x} dy  
\left[-y^2 v_i^2-(1-x-y)^2 v_j^2-2 y \, v_i\cdot v_j (1-x-y)\right]^{n/2-3} 
\, . 
\nonumber
\eeqa
After several manipulations, and with $n=4-\epsilon$, we find
\beqa
I_{1a}&=&\frac{\alpha_s}{\pi} \, (-1)^{-1-\epsilon/2} \, 2^{5\epsilon/2} \, 
\pi^{\epsilon/2} \, \Gamma\left(1+\frac{\epsilon}{2}\right)
(1+\beta^2) \int_0^1 dx \, x^{-1+\epsilon} (1-x)^{-1-\epsilon}
\nonumber \\ && \hspace{-8mm}
\times  \left\{\int_0^1 dz \left[4z \beta^2 (1-z)+1-\beta^2\right]^{-1}
-\frac{\epsilon}{2} \int_0^1 dz \frac{\ln\left[4z \beta^2 (1-z)
+1-\beta^2\right]}{4z \beta^2 (1-z)+1-\beta^2} 
+{\cal O}\left(\epsilon^2\right)\right\} \, .
\nonumber
\eeqa 
The integral over $x$ contains both ultraviolet (UV) and infrared (IR) 
singularities. We isolate the UV singularities via 
\beqa
\int_0^1 dx \, x^{-1+\epsilon} \, (1-x)^{-1-\epsilon}
=\frac{1}{\epsilon}+{\rm IR} \, .
\nonumber
\eeqa
Then the UV pole of the integral is 
\beqa
I_{1a}^{UV}&=&\frac{\alpha_s}{\pi} \frac{(1+\beta^2)}{2\beta}
\frac{1}{\epsilon} \ln\left(\frac{1-\beta}{1+\beta}\right) \, .
\nonumber
\eeqa
Together with the contributions of the top self-energy diagrams, and including color factors, this gives the one-loop result for the cusp anomalous dimension that we presented above.

\subsection{Two-loop cusp anomalous dimension}

\begin{figure}
\centerline{\includegraphics[width=0.5\textwidth]{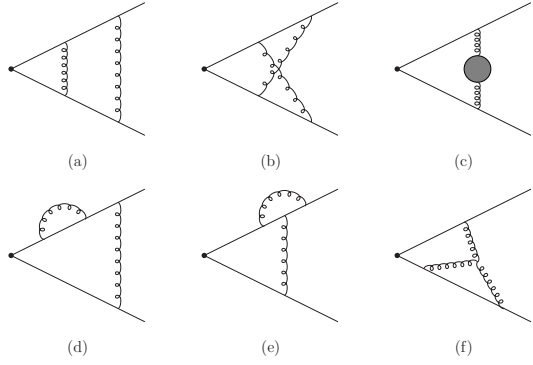}}
\caption{Two-loop vertex-correction diagrams for the cusp anomalous dimension.}
\label{loopg2}
\end{figure}

\begin{figure}
\centerline{\includegraphics[width=0.4\textwidth]{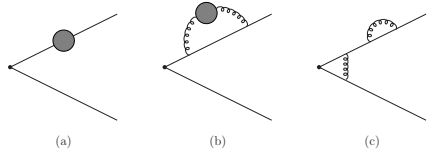}}
\caption{Two-loop top-quark self-energy graphs.}
\label{loopg2s}
\end{figure}

The two-loop vertex-correction graphs for the massive cusp anomalous dimension
are shown in Fig. \ref{loopg2}. Additional two-loop top-quark self-energy graphs that also need to be included are shown in Fig. \ref{loopg2s}.
The grey blobs indicate quark, gluon, and ghost loops.

\begin{figure}
\centerline{\includegraphics[width=0.35\textwidth]{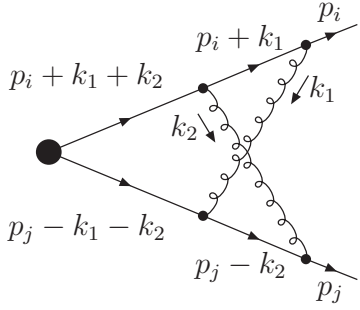}}
\caption{Two-loop crossed diagram.}
\label{loop2cr}
\end{figure}

As an example of the calculation, consider the two-loop crossed diagram
in Fig. \ref{loopg2}(b) with details of momenta assignments in 
Fig. \ref{loop2cr}.
The corresponding integral is 
\beqa
I_{2b}=g_s^4 \int\frac{d^n k_1}{(2\pi)^n}\frac{d^n k_2}{(2\pi)^n}
\frac{(-i)g^{\mu\nu}}{k_1^2} \frac{(-i)g^{\rho\sigma}}{k_2^2}  
\frac{v_i^{\mu}}{v_i\cdot k_1} \frac{v_i^{\rho}}{v_i\cdot (k_1+k_2)}
\frac{(-v_j^{\nu})}{(-v_j\cdot (k_1+k_2))} \frac{(-v_j^{\sigma})}{(-v_j\cdot k_2)}
\, .
\nonumber
\eeqa
We perform the $k_2$ integral first, using Feynman parameterization similarly to the one-loop example, and find 
\beqa
I_{2b}&=& -i \frac{\alpha_s^2}{\pi^2} 2^{-4+\epsilon} \pi^{-2+3\epsilon/2}
\Gamma\left(1-\frac{\epsilon}{2}\right) \Gamma(1+\epsilon) 
(1+\beta^2)^2 \int_0^1 dz 
\nonumber \\ && \hspace{-5mm} \times 
\int_0^1 \frac{dy \, (1-y)^{-\epsilon}}
{\left[2\beta^2(1-y)^2 z^2-2\beta^2(1-y)z-\frac{(1-\beta^2)}{2}\right]^{1-\epsilon/2}} \int \frac{d^nk_1}{k_1^2 \, v_i \cdot k_1 \, [\left((v_i-v_j)z+v_j\right)\cdot k_1]^{1+\epsilon}} \, .
\nonumber
\eeqa
We then proceed with the $k_1$ integral, and isolate the UV and IR poles.
After many steps we find 
\beqa
I_{2b}^{UV}&=&\frac{\alpha_s^2}{\pi^2} \frac{(1+\beta^2)^2}{8\beta^2} 
\frac{1}{\epsilon} \left\{-\frac{1}{3}\ln^3\left(\frac{1-\beta}{1+\beta}\right)
-\ln\left(\frac{1-\beta}{1+\beta}\right)
\left[{\rm Li}_2\left(\frac{(1-\beta)^2}{(1+\beta)^2}\right)
+\zeta_2\right] \right.
\nonumber \\ && \hspace{25mm} \left. 
{}+{\rm Li}_3\left(\frac{(1-\beta)^2}{(1+\beta)^2}\right)-\zeta_3 \right\} \, .
\nonumber
\eeqa

We similarly calculate all other two-loop graphs, and
we include the counterterms for all graphs and multiply with the corresponding
color factors. We determine the two-loop cusp anomalous dimension from the 
UV poles of the sum of the graphs \cite{NKPRL,NK2lproc}:
\beqa
\Gamma_S^{(2)}&=&\frac{K}{2} \, \Gamma_S^{(1)}
+C_F C_A M_{\beta}
\nonumber \\ 
&=&\frac{K}{2} \, \Gamma_S^{(1)}
+C_F C_A \left\{\frac{1}{2}+\frac{\zeta_2}{2}
+\frac{1}{2} \ln^2\left(\frac{1-\beta}{1+\beta}\right) \right.
\nonumber \\ && 
{}-\frac{(1+\beta^2)^2}{8 \beta^2} \left[\zeta_3
+\zeta_2 \ln\left(\frac{1-\beta}{1+\beta}\right)
+\frac{1}{3} \ln^3\left(\frac{1-\beta}{1+\beta}\right) 
{}+\ln\left(\frac{1-\beta}{1+\beta}\right) 
{\rm Li}_2\left(\frac{(1-\beta)^2}{(1+\beta)^2}\right) \right.
\nonumber \\ && \hspace{20mm} \left.
{}-{\rm Li}_3\left(\frac{(1-\beta)^2}{(1+\beta)^2}\right)\right] 
\nonumber \\ && 
{}-\frac{(1+\beta^2)}{4 \beta} \left[\zeta_2
-\zeta_2 \ln\left(\frac{1-\beta}{1+\beta}\right) 
+\ln^2\left(\frac{1-\beta}{1+\beta}\right)
-\frac{1}{3} \ln^3\left(\frac{1-\beta}{1+\beta}\right) \right.
\nonumber \\ && \hspace{20mm} \left. \left.
{}+2  \ln\left(\frac{1-\beta}{1+\beta}\right) 
\ln\left(\frac{(1+\beta)^2}{4 \beta}\right) 
-{\rm Li}_2\left(\frac{(1-\beta)^2}{(1+\beta)^2}\right)\right]\right\}
\nonumber
\eeqa
where, as before,  
$K=C_A (67/18-\zeta_2)-5n_f/9$, and where for shorthand notation and for later 
use we have introduced $M_{\beta}$ to denote all the terms in curly brackets 
in the above equation. As can be seen from the above expression, 
the color structure of $\Gamma_S^{(2)}$ involves only the factors $C_F C_A$ and $C_F n_f$.

In terms of the cusp angle \cite{KorRad}  
$\gamma=\cosh^{-1}(v_i\cdot v_j/\sqrt{v_i^2 v_j^2})=\ln[(1+\beta)/(1-\beta)]$ we can rewrite the one-loop expression as 
$$\Gamma_S^{(1)}=C_F (\gamma \coth\gamma-1)$$  
and the two-loop expression \cite{NKPRL,NK2lproc} as 
\beqa
\Gamma_S^{(2)}&=&\frac{K}{2} \, \Gamma_S^{(1)}
+C_F C_A \left\{\frac{1}{2}+\frac{\zeta_2}{2}+\frac{\gamma^2}{2} \right.
\nonumber \\ && \hspace{25mm} 
{}-\frac{1}{2}\coth^2\gamma\left[\zeta_3-\zeta_2\gamma-\frac{\gamma^3}{3}
-\gamma \, {\rm Li}_2\left(e^{-2\gamma}\right)
-{\rm Li}_3\left(e^{-2\gamma}\right)\right] 
\nonumber \\ && \hspace{25mm} \left.
{}-\frac{1}{2} \coth\gamma\left[\zeta_2+\zeta_2\gamma+\gamma^2
+\frac{\gamma^3}{3}+2\, \gamma \, \ln\left(1-e^{-2\gamma}\right)
-{\rm Li}_2\left(e^{-2\gamma}\right)\right] \right\}.
\nonumber
\eeqa

The cusp anomalous dimension is an essential component of other calculations
for QCD processes, where the 
color structure gets more complicated with more than two colored partons 
in the process. 

\begin{figure}
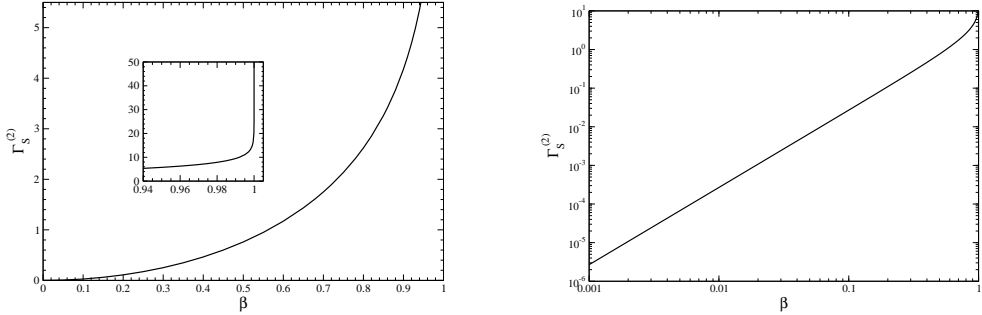

\centerline{\includegraphics[width=0.4\textwidth]{Gammainsetplot.eps}
\hspace{10mm}
\includegraphics[width=0.4\textwidth]{Gamma2logplot.eps}}
\caption{The two-loop cusp anomalous dimension, $\Gamma_S^{(2)}$, as a function of $\beta$ in a linear (left) and logarithmic (right) plot.}
\label{Gammaplots}
\end{figure}

Linear and logarithmic plots of $\Gamma_S^{(2)}$ are shown in 
Fig. \ref{Gammaplots}.
$\Gamma_S^{(2)}$ vanishes at $\beta=0$, the threshold limit, and diverges 
at $\beta=1$, the massless limit. 

We next determine analytically the small and large $\beta$ behavior of 
$\Gamma_S^{(2)}$.
For the small $\beta$ behavior we expand around $\beta=0$ and find 
\beqa
\Gamma_{S \, {\rm exp}}^{(2)}&=&-\frac{2}{27} \beta^2 
\left[C_F C_A (18 \zeta_2-47)+5 C_F n_f\right]+{\cal O}(\beta^4) \, .
\nonumber
\eeqa
We note that $\Gamma_S^{(2)}$ is an even function of $\beta$. 
For the large $\beta$ behavior, as $\beta \rightarrow 1$, we find 
$\Gamma_S^{(2)} \rightarrow \frac{K}{2} \Gamma_S^{(1)}$.

We next construct an approximation valid for all $\beta$ \cite{NKPRL}:
\beqa
\Gamma^{(2)}_{S \, {\rm approx}}&=&\Gamma^{(2)}_{S \, {\rm exp}}
+\frac{K}{2} \Gamma_S^{(1)}-\frac{K}{2} \Gamma^{(1)}_{S \, {\rm exp}}
\nonumber \\ &=&
\frac{K}{2} \Gamma_S^{(1)}+C_F C_A \left(1-\frac{2}{3}\zeta_2\right) \beta^2
+{\cal O}\left(\beta^4\right) \, .
\nonumber
\eeqa

\begin{figure}
\centerline{\includegraphics[width=0.55\textwidth]{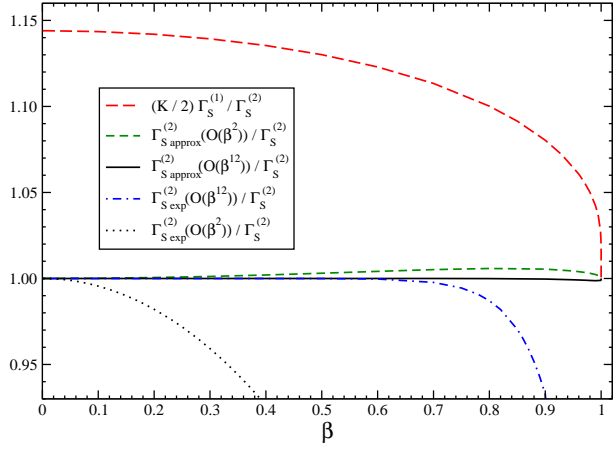}}
\caption{Expansions and approximations for $\Gamma_S^{(2)}$.}
\label{Gammaapproxplot}
\end{figure}

The expansions and approximations to $\Gamma_S^{(2)}$ are shown in 
Fig. \ref{Gammaapproxplot}. 
$\Gamma^{(2)}_{S \, {\rm approx}}$ is a remarkably good approximation to the complete 
$\Gamma_S^{(2)}$.

\section{Soft anomalous dimension matrices for $t{\bar t}$ production}

\begin{figure}
\centerline{\includegraphics[width=0.2\textwidth]{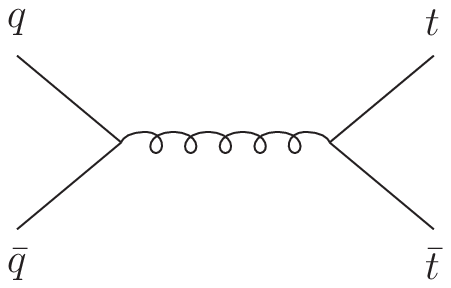}
\hspace{10mm}\includegraphics[width=0.7\textwidth]{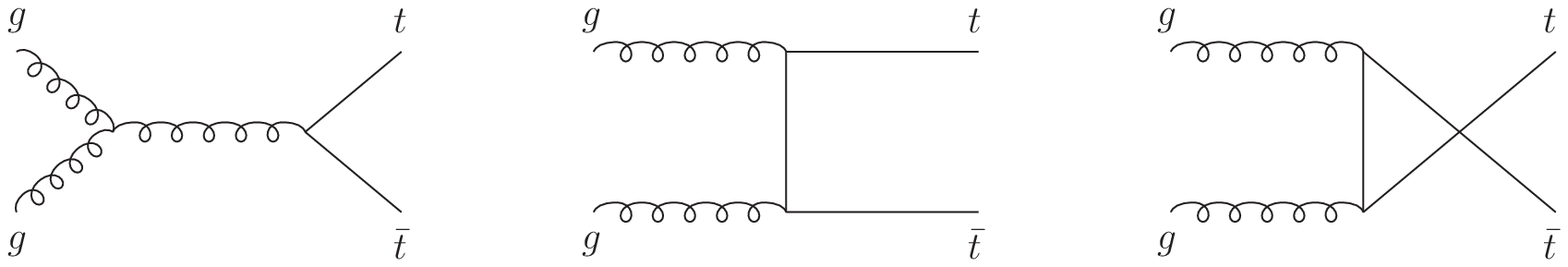}}
\caption{Lowest-order diagrams for the $q{\bar q} \rightarrow t{\bar t}$ 
channel (left diagram) and the $gg \rightarrow t{\bar t}$ channel 
(right three diagrams).}
\label{qqggdiag}
\end{figure}

The top-antitop pair production partonic processes at LO are

$$q(p_1)+{\bar q}(p_2) \rightarrow t(p_3)+ {\bar t}(p_4)$$

and

$$g(p_1)+g(p_2) \rightarrow t(p_3)+{\bar t}(p_4)$$

The LO diagrams for these processes are shown in Fig. \ref{qqggdiag}.
We define $s=(p_1+p_2)^2$,  $t_1=(p_1-p_3)^2-m_t^2$, $u_1=(p_2-p_3)^2-m_t^2$,
and $\beta=\sqrt{1-4m_t^2/s}$. Note that $\beta$ is the top-quark speed in 
the LO kinematics. At the Tevatron and the LHC the $t{\bar t}$ cross section 
receives most contributions in the region around $0.3 < \beta < 0.8$ 
which peak roughly around $\beta \sim 0.6$. 

We next present the results at one and two loops for the soft anomalous matrices for these partonic processes. 
The soft anomalous dimension matrix for $q(p_1)+{\bar q}(p_2) \rightarrow t(p_3)+{\bar t}(p_4)$ in a color tensor basis
consisting of singlet and octet exchange in the $s$ channel,
\beqa
c_1 = \delta_{12}\delta_{34}\, , \quad \quad
c_2 =  T^c_{F\; 21} \, T^c_{F\; 34}\, ,
\nonumber
\eeqa
has elements
\beqa
\Gamma_S^{ q{\bar q}\rightarrow t{\bar t}}=\left[\begin{array}{cc}
\Gamma_{q{\bar q} \, 11} & \Gamma_{q{\bar q} \, 12} \\
\Gamma_{q{\bar q} \, 21} & \Gamma_{q{\bar q} \, 22}
\end{array}
\right] \, .
\nonumber
\eeqa

At one loop we find \cite{NKGS,NKtop}
\beqa
\Gamma_{q{\bar q} \,11}^{(1)}&=&-C_F \, [L_{\beta}+1]
\nonumber \\
\Gamma_{q{\bar q} \,12}^{(1)}&=&
\frac{C_F}{C_A} \ln\left(\frac{t_1}{u_1}\right) 
\nonumber \\ 
\Gamma_{q{\bar q} \,21}^{(1)}&=&
2\ln\left(\frac{t_1}{u_1}\right) \hspace{15mm}
\nonumber \\ 
\Gamma_{q{\bar q} \,22}^{(1)}&=&C_F
\left[4\ln\left(\frac{t_1}{u_1}\right)
-L_{\beta}-1\right]
+\frac{C_A}{2}\left[-3\ln\left(\frac{t_1}{u_1}\right)
+\ln\left(\frac{t_1u_1}{s m_t^2}\right)+L_{\beta}\right]
\nonumber
\eeqa
where
$L_{\beta}=\frac{1+\beta^2}{2\beta}\ln\left(\frac{1-\beta}{1+\beta}\right)$. 
We note that the first element of this matrix is identical to the one-loop massive cusp anomalous dimension.

Then the elements of the soft anomalous dimension matrix for 
the process $q{\bar q} \rightarrow t{\bar t}$ at two loops are \cite{NKtop}
\beqa
\Gamma_{q{\bar q} \,11}^{(2)}&=&\frac{K}{2} \Gamma_{q{\bar q} \,11}^{(1)}
+C_F C_A \, M_{\beta}
\nonumber \\ 
\Gamma_{q{\bar q} \,12}^{(2)}&=&
\frac{K}{2} \Gamma_{q{\bar q} \,12}^{(1)} -\frac{C_F}{2} N_{\beta} \ln\left(\frac{t_1}{u_1}\right) 
\nonumber \\
\Gamma_{q{\bar q} \,21}^{(2)}&=&
\frac{K}{2}  \Gamma_{q{\bar q} \,21}^{(1)} +C_A N_{\beta} \ln\left(\frac{t_1}{u_1}\right) 
\nonumber \\
\Gamma_{q{\bar q} \,22}^{(2)}&=&
\frac{K}{2} \Gamma_{q{\bar q} \,22}^{(1)}
+C_A\left(C_F-\frac{C_A}{2}\right) \, M_{\beta} \nonumber
\eeqa
We note that the first element of this matrix is identical to the two-loop massive cusp anomalous dimension, and $M_{\beta}$ was defined in the previous section.
Here $N_{\beta}$ is a subset of the terms of $M_{\beta}$, 
\beqa
N_{\beta}&=&-\frac{(1+\beta^2)}{4 \beta} \left[
\ln^2\left(\frac{1-\beta}{1+\beta}\right)
+2  \ln\left(\frac{1-\beta}{1+\beta}\right) 
\ln\left(\frac{(1+\beta)^2}{4 \beta}\right) 
-{\rm Li}_2\left(\frac{(1-\beta)^2}{(1+\beta)^2}\right)\right]
\nonumber \\ && 
{}+\frac{1}{2}\ln^2\left(\frac{1-\beta}{1+\beta}\right) \, .
\nonumber
\eeqa

The soft anomalous dimension matrix for $g(p_1)+g(p_2) \rightarrow t(p_3)+{\bar t}(p_4)$ in a color tensor basis
\beqa
c_1=\delta^{12}\,\delta_{34}, \quad c_2=d^{12c}\,T^c_{34},
\quad c_3=i f^{12c}\,T^c_{34} 
\nonumber
\eeqa
where $d$ and $f$ are the totally symmetric and 
antisymmetric $SU(3)$ invariant tensors, is
\beqa
\Gamma_S^{gg\rightarrow t{\bar t}}=\left[\begin{array}{ccc}
\Gamma_{gg\, 11} & 0 & \Gamma_{gg\,13} \vspace{2mm} \\
0 & \Gamma_{gg\,22} & \Gamma_{gg\,23} \vspace{2mm} \\
\Gamma_{gg\,31} & \Gamma_{gg\,32} & \Gamma_{gg\,22}
\end{array}
\right] \, .
\nonumber
\eeqa

At one loop we have \cite{NKGS,NKtop}
\beqa
\Gamma_{gg\, 11}^{(1)}&=&-C_F [L_{\beta}+1]
\nonumber \\
\Gamma_{gg\, 13}^{(1)}&=& \ln\left(\frac{t_1}{u_1}\right) 
\nonumber \\
\Gamma_{gg\, 31}^{(1)}&=&2 \ln\left(\frac{t_1}{u_1}\right) 
\nonumber \\
\Gamma_{gg\, 22}^{(1)}&=&-C_F [L_{\beta}+1] 
+\frac{C_A}{2}\left[\ln\left(\frac{t_1 u_1}{m^2 s}\right)+L_{\beta}\right] 
\nonumber \\
\Gamma_{gg\, 23}^{(1)}&=&\frac{C_A}{2} \ln\left(\frac{t_1}{u_1}\right) 
\nonumber \\
\Gamma_{gg\, 32}^{(1)}&=&\frac{N_c^2-4}{2N_c} \ln\left(\frac{t_1}{u_1}\right) 
\nonumber
\eeqa

At two loops we find \cite{NKtop}
\beqa
\Gamma_{gg\, 11}^{(2)}&=& \frac{K}{2} \Gamma_{gg \,11}^{(1)}
+C_F C_A \, M_{\beta}
\nonumber \\
\Gamma_{gg\, 13}^{(2)}&=&\frac{K}{2} \Gamma_{gg \,13}^{(1)} 
-\frac{C_A}{2} N_{\beta} \ln\left(\frac{t_1}{u_1}\right) 
\nonumber \\
\Gamma_{gg\, 31}^{(2)}&=&\frac{K}{2} \Gamma_{gg \,31}^{(1)} 
+C_A N_{\beta} \ln\left(\frac{t_1}{u_1}\right) 
\nonumber \\
\Gamma_{gg\, 22}^{(2)}&=& \frac{K}{2} \Gamma_{gg \,22}^{(1)}
+C_A \left(C_F-\frac{C_A}{2}\right) \, M_{\beta} 
\nonumber \\
\Gamma_{gg\, 23}^{(2)}&=&\frac{K}{2} \Gamma_{gg \,23}^{(1)} 
\nonumber \\
\Gamma_{gg\, 32}^{(2)}&=&\frac{K}{2} \Gamma_{gg \,32}^{(1)}  
\nonumber
\eeqa

\section{Double-differential kinematics}

We consider a generic hadronic process with momenta $p_{h_1}+p_{h_2} \rightarrow p_3+p_4$ with underlying partonic process $p_1+p_2 \rightarrow p_3+p_4$.
We write general kinematics formulas that can be used for both top-antitop pair 
and single-top production.

\subsection{Kinematics with $S$, $T$, $U$}

The hadronic variables are 
$S=(p_{h1}+p_{h2})^2$, $T=(p_{h1}-p_3)^2$, $U=(p_{h2}-p_3)^2$.
The partonic variables are 
$s=(p_1+p_2)^2$, $t=(p_1-p_3)^2$, $u=(p_2-p_3)^2$; we also define 
$s_4=s+t+u-m_3^2-m_4^2$ which describes the excess energy for additional radiation in the process, and thus measures kinematical distance from partonic threshold. Note that $p_1=x_1 p_{h1}$, $p_2=x_2 p_{h2}$, $s=x_1 x_2 S$, $t-m_3^2=x_1(T-m_3^2)$, $u-m_3^2=x_2(U-m_3^2)$, with $x_1$ and $x_2$ the momentum fractions of the colliding partons in the corresponding hadrons.
The total hadronic cross section is found by integrating over the double-differential partonic cross section convoluted with the parton distribution functions $\phi$:
\beqa
\sigma_{p_{h1}p_{h2} \rightarrow p_3 p_4}(S)&=&
\int_{T_{min}}^{T_{max}} dT 
\int_{U_{min}}^{U_{max}} dU 
\int_{x_{2min}}^1 dx_2 
\int_0^{s_{4max}} ds_4
\nonumber \\ &&
\times \frac{x_1 x_2}{x_2 S+T-m_3^2} \,
\phi(x_1) \, \phi(x_2) \, 
\frac{d^2{\hat\sigma}_{p_1 p_2 \rightarrow p_3 p_4}}{dt \, du}
\nonumber
\eeqa
where
\beqa
x_1=\frac{s_4-m_3^2+m_4^2-x_2(U-m_3^2)}{x_2 S+T-m_3^2}
\nonumber
\eeqa
\beqa
T_{^{max}_{min}}=-\frac{1}{2}(S-m_3^2-m_4^2) \pm 
\frac{1}{2} \sqrt{(S-m_3^2-m_4^2)^2-4m_3^2m_4^2} 
\nonumber
\eeqa
\beqa
U_{max}=m_3^2+\frac{S \, m_3^2}{T-m_3^2}
\nonumber
\eeqa
$U_{min}=-S-T+m_3^2+m_4^2$,
$x_{2min}=(m_4^2-T)/(S+U-m_3^2)$
and $s_{4max}=x_2(S+U-m_3^2)+T-m_4^2$.

\subsection{Kinematics with $p_T$ and rapidity}

We next provide an alternative cross-section calculation in terms of the transverse momentum, $p_T$, and the rapidity, $Y$, of the outgoing particle with momentum $p_3$. 
We further define $T_1=T-m_3^2$, $U_1=U-m_3^2$, $t_1=t-m_3^2$, and $u_1=u-m_3^2$.
Also, 
$U_1=-\sqrt{S} m_T e^Y$ and $T_1=-\sqrt{S} m_T e^{-Y}$
with $m_T=\sqrt{m_3^2+p_T^2}$. We then calculate the total hadronic cross section via
\beqa
\sigma_{p_{h1}p_{h2} \rightarrow p_3 p_4}(S)&=&
\int_0^{p_{T \,max}^2} dp_T^2 
\int_{Y^-}^{Y^+} dY 
\int_{x_1^-}^1 dx_1 
\int_0^{s_{4max}} ds_4
\nonumber \\ &&
\times \frac{x_1 x_2 \, S}{x_1 S+U_1} \,
\phi(x_1) \, \phi(x_2) \, 
\frac{d^2{\hat\sigma}_{p_1 p_2 \rightarrow p_3 p_4}}{dt_1 \, du_1}
\nonumber
\eeqa
where
\beqa
x_2=\frac{s_4-m_3^2+m_4^2-x_1T_1}{x_1 S+U_1},
\nonumber
\eeqa
\beqa
p_{T \, max}^2=\frac{(S-m_3^2-m_4^2)^2-4m_3^2 m_4^2}{4S}
\nonumber
\eeqa
\beqa
Y^{\pm}=\pm \frac{1}{2} \ln \frac{1+\sqrt{1-\frac{4m_T^2}
{S[1+(m_3^2-m_4^2)/S]^2}}}
{1-\sqrt{1-\frac{4m_T^2}{S[1+(m_3^2-m_4^2)/S]^2}}}
\nonumber
\eeqa
\beqa
x_1^-=\frac{-(U_1+m_3^2-m_4^2)}{S+T_1}
\nonumber
\eeqa
\beqa
s_{4max}=x_1(S+T_1)+U_1+m_3^2-m_4^2 \, .
\nonumber
\eeqa

\section{Total cross section for $t{\bar t}$ production}

\begin{figure}
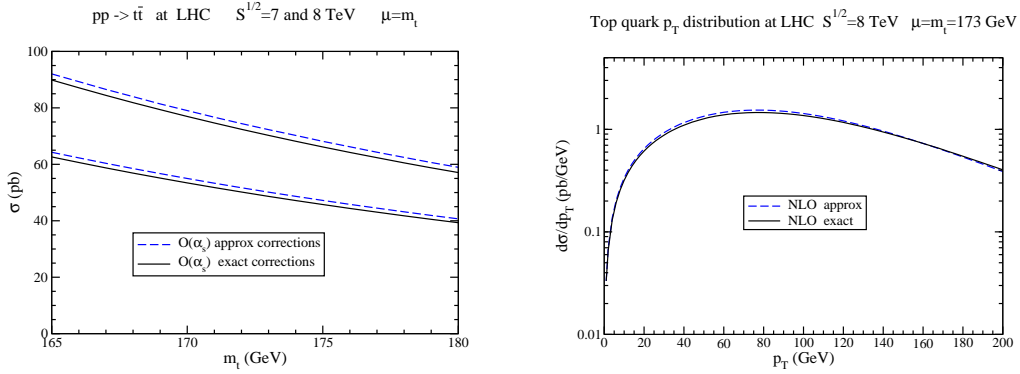

\centerline{\includegraphics[width=0.42\textwidth]{top1a1elhcplot.eps}
\hspace{10mm}\includegraphics[width=0.42\textwidth]{ptcorr8lhcmplot.eps}}
\caption{(Left) NLO exact and approximate corrections to the top-pair cross section at 7 TeV (lower lines) and 8 TeV (upper lines) LHC energy; (Right) NLO exact and approximate top-quark transverse momentum distributions at 8 TeV.}
\label{topptcorr}
\end{figure}

\begin{figure}
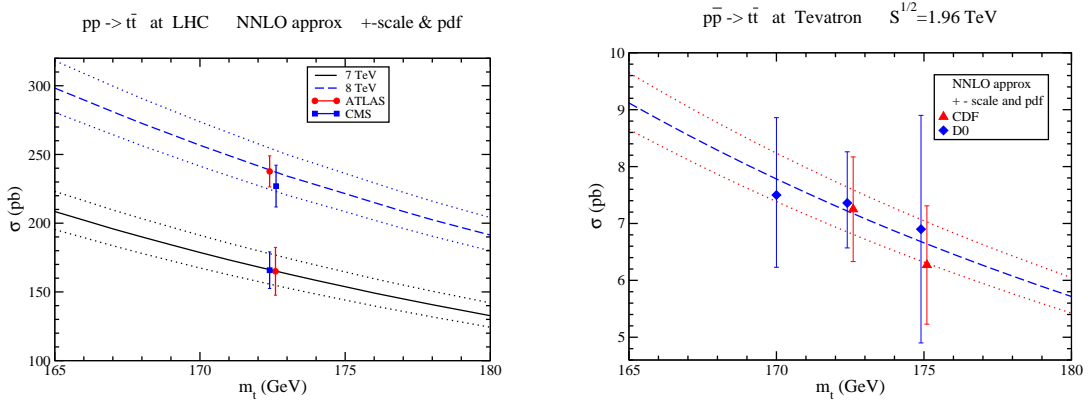

\centerline{\includegraphics[width=0.45\textwidth]{ttblhcplot.eps}
\hspace{10mm}\includegraphics[width=0.45\textwidth]{ttbtevplot.eps}}
\caption{The top-pair total cross section at LHC (left) and Tevatron (right) energies.}
\label{ttbar}
\end{figure}

We begin our presentation of numerical results with the total cross section 
for top-antitop pair production. We use the MSTW2008 NNLO  \cite{MSTW} 
parton distribution functions (pdf) for all the numerical results.
We first show that the threshold approximation works very well both for total 
cross sections and differential distributions. 

We denote the NLO soft-gluon corrections from the expansion of the NNLL resummed
cross section as NLO approximate corrections. Similarly the NNLO soft-gluon 
corrections are denoted as NNLO approximate corrections. Furthermore, the sum 
of the exact NLO cross section and the NNLO approximate corrections is denoted as the NNLO approximate cross section (and this applies to both total and differential cross sections).

Figure \ref{topptcorr} shows that the NLO exact and approximate corrections to the total cross section as well as the top-quark $p_T$ distribution are nearly identical. We have an excellent approximation: there is 
less than 1\% difference between NLO approximate and exact cross sections.
For the best prediction we add the NNLO approximate corrections to the exact NLO cross section. We find that that the scale dependence is greatly reduced when the NNLO approximate corrections are included.

In Fig. \ref{ttbar} we display theoretical predictions at approximate NNLO for the total cross section as a function of top-quark mass at the LHC (left plot) and the Tevatron (right plot) and compare them with data from the LHC at 7 TeV \cite{ATLAS7ttbar,CMS7ttbar} 
and 8 TeV \cite{ATLAS8ttbar,CMS8ttbar} and from the Tevatron at 1.96 TeV \cite{CDFttbar,D0ttbar}. We find very good agreement between the theoretical predictions and the data. The approximate NNLO prediction \cite{NKtop} for $m_t=173$ GeV 
is $7.08 {}^{+0.20}_{-0.24} {}^{+0.36}_{-0.27}$ pb at the Tevatron at 1.96 TeV; 
$163 {}^{+7}_{-5} \pm 9$ pb at the LHC at 7 TeV; $234 {}^{+10}_{-7} \pm 12$ pb at 
8 TeV LHC; and $920 {}^{+50}_{-39} {}^{+33}_{-35}$ pb at 14 TeV LHC.
The central result is with $\mu_F=\mu_R=m_t$, the first uncertainty is from independent variation of $\mu_F$ and $\mu_R$ over the range $m_t/2$ to $2m_t$, and the second uncertainty is from the MSTW2008 NNLO pdf at 90\% CL. Of course the numerical results depend on the choice of pdf, $\alpha_s$, and choices of top quark mass and scales. 

There are many differences between various resummation/NNLO approximate approaches in the literature and these have been detailed previously in \cite{NKBP,pnsnoweps}. The differences include whether the resummation is for the total-only cross section versus for the double-differential cross section, whether it uses moment-space perturbative QCD (pQCD) versus Soft-Collinear Effective Theory (SCET), etc.

\begin{figure}
\centerline{\includegraphics[width=0.5\textwidth]{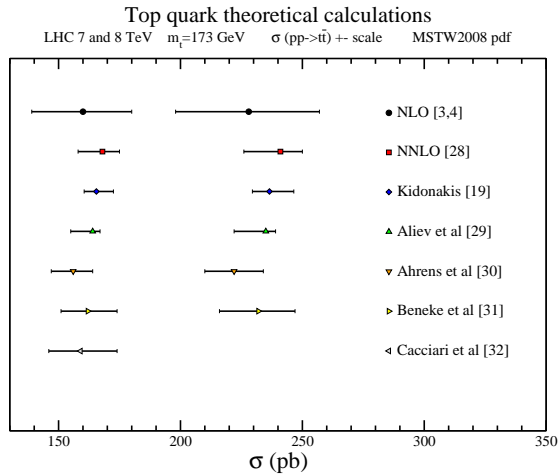}}
\caption{Theoretical results for the $t{\bar t}$ cross section at 7 and 8 TeV LHC energies.}
\label{theorylhcplot}
\end{figure}

Resummations that only use the total cross section refer to production (or absolute) threshold and resum logarithms of $\beta = \sqrt{1-4m_t^2/s}$. The soft limit here is the production threshold limit $\beta \rightarrow 0$ (where the top quark velocities are zero), which is a special case of the more general partonic threshold. The partonic threshold is where there is just enough energy to produce the top quarks but they can have arbitrary $p_T$ and are not restrained to be at rest. This more general double-differential approach to resummation can be expressed in single-particle-inclusive kinematics for the differential cross section $d\sigma/dp_T dy$ where the logarithms involve $s_4= s+t_1+u_1$ and the soft limit is $s_4 \rightarrow 0$. 

The double-differential approach, in addition to using a more general definition of threshold, also allows the calculation of transverse momentum and rapidity distributions. For differential calculations, further differences between approaches arise from how the relation $s+t_1+u_1=0$ is used in the plus-distribution coefficients, how subleading terms are treated, if/how damping factors are implemented to reduce the influence of contributions far from threshold, etc.  

A comparison of various NNLO approximate approaches is shown in Fig. \ref{theorylhcplot}, all with the same choice of parameters, at 7 and 8 TeV LHC energies.
In addition, exact NLO \cite{NLOtt1,NLOtt2} and NNLO \cite{NNLO} results for the total cross sections are also shown on the plot.

Ref. \cite{NKtop} uses our pQCD resummation formalism for the double-differential cross section. Ref. \cite{ALLMUW} uses pQCD resummation for the total-only cross section. Ref. \cite{AFNPY} uses the SCET resummation formalism for the double-differential cross section. Ref. \cite{BFKS} uses the SCET resummation formalism for the total-only cross section. Lastly, Ref. \cite{CCMMN}  uses pQCD resummation for the total-only cross section. 

\begin{figure}
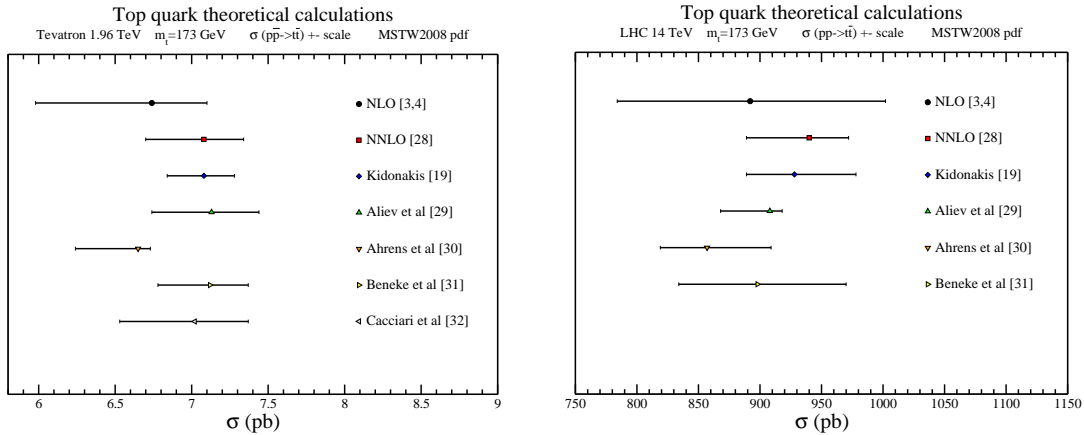

\centerline{\includegraphics[width=0.45\textwidth]{theorytevplot.eps}
\hspace{7mm}\includegraphics[width=0.47\textwidth]{theory14lhcplot.eps}}
\caption{Theoretical results for the $t{\bar t}$ cross section at the Tevatron (left) and at 14 TeV LHC energy (right).}
\label{theorytev14}
\end{figure}

Figure \ref{theorytev14} shows the corresponding comparison for the Tevatron (left) and for 14 TeV LHC energy (right).
One notes the varying degree of success of the various approaches in approximating the exact NNLO result.

The result in Ref. \cite{NKtop} from our formalism is very close to the exact NNLO \cite{NNLO} result: both the central values and the scale uncertainty are nearly the same and this holds true for all collider energies and top quark masses.
This was expected from the comparison of exact and approximate corrections at NLO for both total and differential cross sections, and also from the comparison of approximate NNLO results in different kinematics in 2003 \cite{PRD68} (see also the discussions in \cite{NKtop} and \cite{pnsnoweps}).
There is less than 1\% difference between approximate and exact cross sections at both NLO and NNLO.

The stability of the theoretical prediction over the past decade and the reliability of the NNLO approximate result and near-identical value to 
exact NNLO is very important for several reasons:

$\bullet$ it provides confidence for applications to other processes (notably single-top production, $W$ production, and other processes);

$\bullet$ it has been used extensively as a background for many analyses (Higgs searches, etc);

$\bullet$ it means that we have reliable and near-exact NNLO $p_T$ and rapidity distributions, which is important since at present there do not exist any exact NNLO results for differential distributions;

$\bullet$ it suggests that NNNLO soft-gluon corrections may be good approximations to exact results at that order if/when they ever become available.

\section{Top-quark $p_T$ and rapidity distributions in $t{\bar t}$ production}

We continue with top quark differential distributions in $t{\bar t}$ 
production.
We present theoretical results for the top-quark transverse momentum and 
rapidity distributions at Tevatron and LHC energies.

\subsection{Top-quark $p_T$ distribution}

\begin{figure}
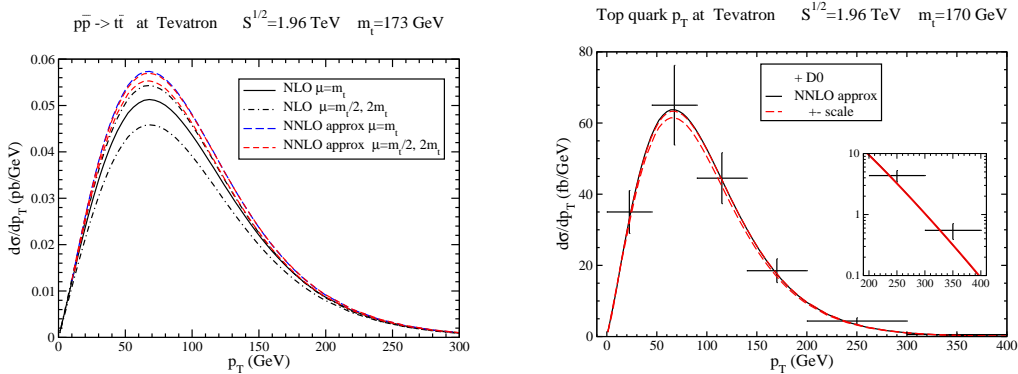

\centerline{\includegraphics[width=0.42\textwidth]{pttev1plot.eps}
\hspace{10mm}\includegraphics[width=0.42\textwidth]{ptD0tevplot.eps}}
\caption{The top quark $p_T$ distribution at the Tevatron.}
\label{pttev}
\end{figure}

\begin{figure}
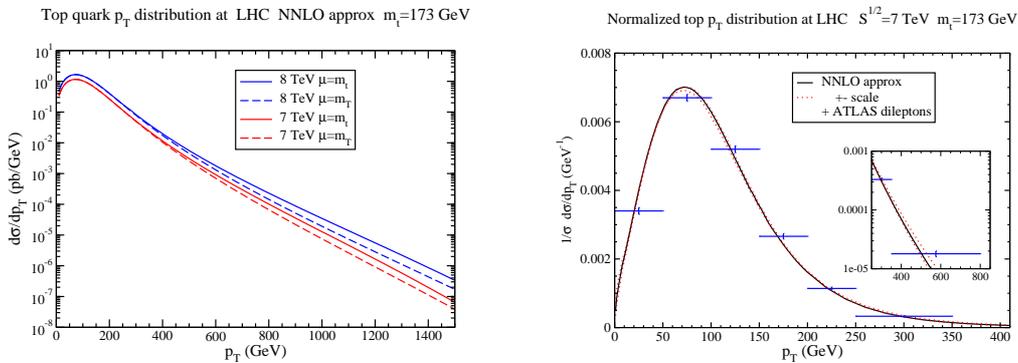

\centerline{\includegraphics[width=0.42\textwidth]{ptlhcmTmplot.eps}
\hspace{10mm}\includegraphics[width=0.42\textwidth]{pt7lhcnormATLASplot.eps}}
\caption{(Left) The top quark $p_T$ distribution at the LHC at 7 and 8 TeV energies. (Right) The normalized top quark $p_T$ distribution at 7 TeV LHC energy compared with ATLAS data in the dileptons channel.}
\label{ptlhc}
\end{figure}

Figure \ref{pttev} displays the theoretical top quark $p_T$ distribution at the Tevatron. A reduction in scale dependence relative to NLO is observed when the NNLO soft-gluon corrections are included.
Excellent agreement of the NNLO approximate results with D0 data \cite{D0pt} can be seen  over all the $p_T$ range from the plot on the right. The theoretical results are also in good agreement with newer D0 data \cite{D0pty2013}.

The top quark $p_T$ distribution at LHC energies is shown on the left plot of Fig. \ref{ptlhc} with two different choices of scale: $m_t$ and $m_T=\sqrt{p_T^2+m_t^2}$. The right plot shows a comparison of the theoretical approximate NNLO normalized $p_T$ distribution, $(1/\sigma) d\sigma/dp_T$, to recent ATLAS data \cite{ATLAS7pty} at 7 TeV energy up to $p_T$ of 800 GeV. 

\begin{figure}
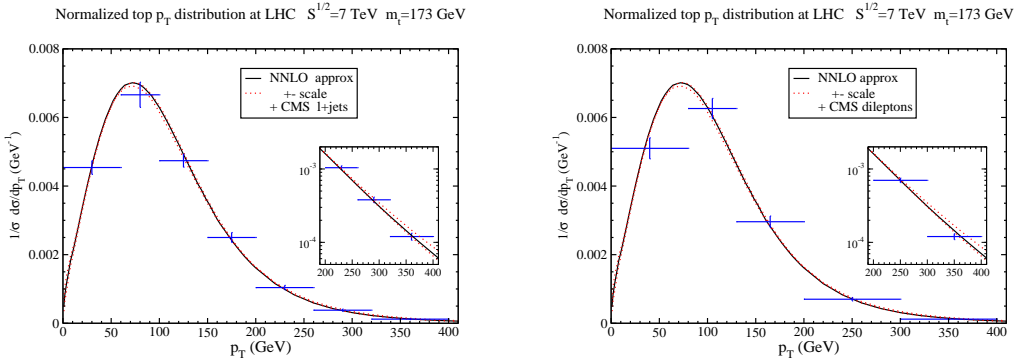

\centerline{\includegraphics[width=0.42\textwidth]{pt7lhcnormCMSljetplot.eps}
\hspace{10mm}\includegraphics[width=0.42\textwidth]{pt7lhcnormCMSdileptplot.eps}}
\caption{The normalized top quark $p_T$ distribution at 7 TeV LHC energy compared with CMS data in the $\ell$+jets channel (left) and the dileptons channel (right).}
\label{ptnorm}
\end{figure}

Figure \ref{ptnorm} shows the same theoretical normalized top quark $p_T$ distribution at the LHC compared to CMS data \cite{CMS7pty} at 7 TeV energy in the $\ell$+jets channel (left plot) and the dileptons channel (right plot). There is excellent agreement with CMS data, and the NNLO approximate result describes the data better than event generators (see the discussion in \cite{CMS7pty}); a similar conclusion is drawn in the comparison with CMS data at 8 TeV energy \cite{CMS8pty}.

\subsection{Top-quark rapidity distribution}

\begin{figure}
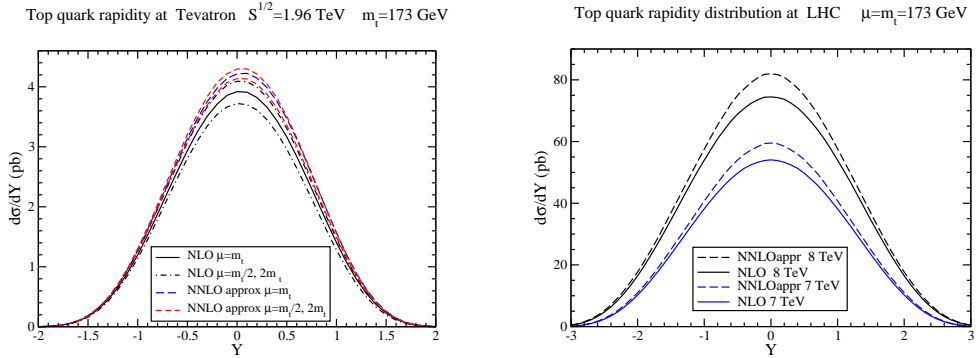

\centerline{\includegraphics[width=0.4\textwidth]{ytevplot.eps}
\hspace{10mm}\includegraphics[width=0.4\textwidth]{ylhcmplot.eps}}
\caption{The top quark rapidity distribution at the Tevatron (left) and the LHC (right).}
\label{ytevlhc}
\end{figure}

The top-quark rapidity distribution has been calculated \cite{NKtopy} 
for Tevatron energy (left plot in Fig. \ref{ytevlhc})
and is in good agreement with recent data from D0 \cite{D0pty2013}.

The top-quark forward-backward asymmetry is defined by 
$$A_{\rm FB}=\frac{\sigma(Y>0)-\sigma(Y<0)}{\sigma(Y>0)+\sigma(Y<0)} \, .$$
The asymmetry is significant at the Tevatron.
The theoretical result \cite{NKtopy} for Tevatron energy is 
$A_{\rm FB}=0.052^{+0.000}_{-0.006}$
which is significantly smaller than observed values.

The theoretical top quark rapidity distribution at LHC energies \cite{NKtopy} 
is shown in the right plot of Fig. \ref{ytevlhc} at NLO and approximate NNLO.

\begin{figure}
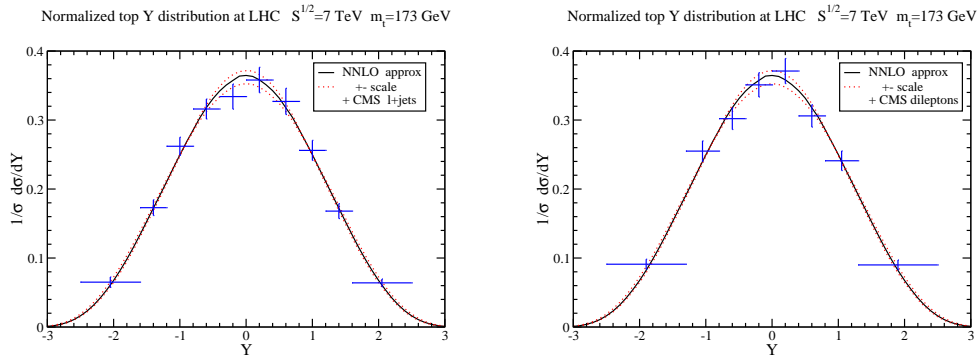

\centerline{\includegraphics[width=0.4\textwidth]{y7lhcnormCMSljetplot.eps}
\hspace{10mm}\includegraphics[width=0.4\textwidth]{y7lhcnormCMSdileptplot.eps}}
\caption{The normalized top quark rapidity distribution at 7 TeV LHC energy compared with CMS data in the $\ell$+jets channel (left) and the dileptons channel (right).}
\label{ynorm}
\end{figure}

The normalized top quark rapidity distribution at the LHC at 7 TeV energy is shown in Fig. \ref{ynorm}. Excellent agreement is found with CMS data at 7 TeV \cite{CMS7pty} and also
at 8 TeV \cite{CMS8pty}.

\section{Single-top production}

\begin{figure}
\centerline{\includegraphics[width=0.2\textwidth]{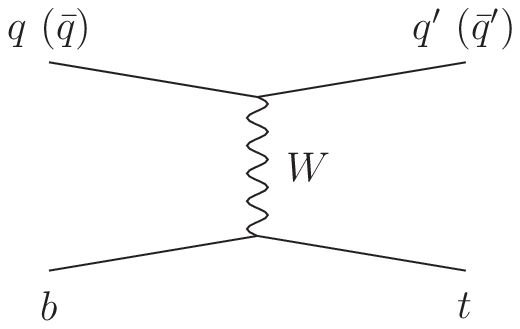}
\hspace{7mm} \includegraphics[width=0.2\textwidth]{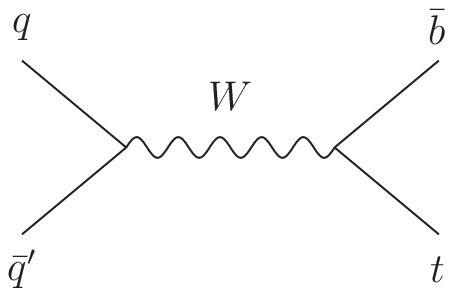}
\hspace{7mm} \includegraphics[width=0.45\textwidth]{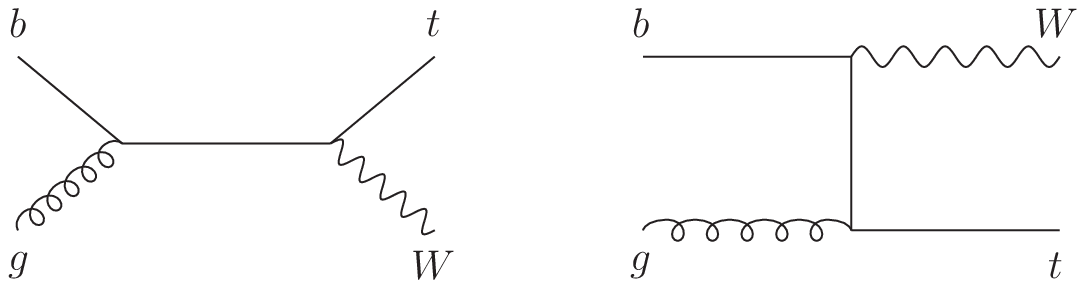}}
\caption{LO diagrams for single-top production in the $t$-channel (left diagram), $s$-channel (second from left), and in $tW$ production (right two diagrams).}
\label{singletopdiag}
\end{figure}

Single-top-quark production was first observed at the Tevatron in 2009 \cite{D0st,CDFst}. The single-top partonic processes at LO are shown in Fig. \ref{singletopdiag}.

The $t$-channel processes are of the form  $qb \rightarrow q' t$ and ${\bar q} b \rightarrow {\bar q}' t$ and are numerically dominant at Tevatron and LHC energies.
The $s$-channel processes are of the form $q{\bar q}' \rightarrow {\bar b} t$
and are small at both the Tevatron and the LHC.
The associated $tW$ production proceeds via $bg \rightarrow tW^-$
and is negligible at the Tevatron but significant (second largest) at the LHC.
A related process to $tW$ production is the associated production of a charged Higgs boson with a top quark, $bg \rightarrow tH^-$.

\subsection{$t$-channel production}

We begin with single top quark production in the $t$-channel. 
This is the dominant single-top production channel at both Tevatron and LHC energies. The complete NLO corrections were calculated in \cite{BWHL}.

The soft anomalous dimension matrix for $t$-channel single top production at one and two loops has been calculated \cite{NKsingletopTev,NKtch}. The first element of this $2\times 2$ matrix is given at one-loop by \cite{NKsingletopTev,NKtch}
\beqa
{\Gamma}_{S\, t-11}^{(1)}
=C_F \left[\ln\left(\frac{-t}{s}\right)
+\ln\left(\frac{m_t^2-t}{m_t\sqrt{s}}\right)-\frac{1}{2}\right] 
\nonumber
\eeqa
and at two loops by \cite{NKtch}
\beqa
\Gamma_{S\, t-11}^{(2)}=\frac{K}{2}\Gamma_{S\, t-11}^{(1)}
+C_F C_A \frac{(1-\zeta_3)}{4} \, .
\nonumber
\eeqa

\begin{figure}
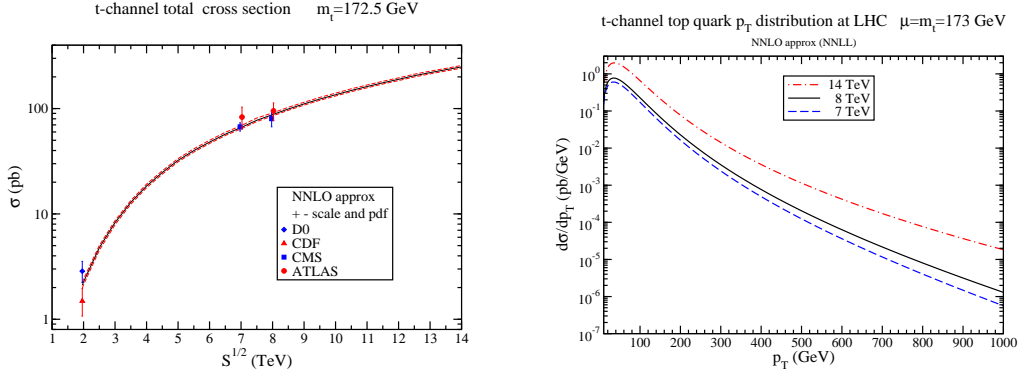

\centerline{\includegraphics[width=0.42\textwidth]{tchtotalSlhcplot.eps}
\hspace{10mm}\includegraphics[width=0.42\textwidth]{pttchtoplhclogplot.eps}}
\caption{The $t$-channel total cross section (left); the top-quark $p_T$ distribution in $t$-channel production (right).}
\label{tchtop}
\end{figure}

The left plot of Fig. \ref{tchtop} shows the total $t$-channel cross section as a function of collider energy. Excellent agreement is found with 
D0 \cite{D0tands}, CDF \cite{CDFtands}, CMS \cite{CMStch7,CMStch8}, and ATLAS \cite{ATLAStch7,ATLAStch8} results.  

\begin{table}
\centerline{\begin{tabular}{c|c|c|c}
LHC  & $t$ &  
${\bar t}$ & Total (pb) \\ 
\hline
7 TeV  &  $43.0 {}^{+1.6}_{-0.2} \pm 0.8$
&  $22.9 \pm 0.5 {}^{+0.7}_{-0.9}$
&  $65.9 {}^{+2.1}_{-0.7} {}^{+1.5}_{-1.7}$
\\
 8 TeV  &  $56.4 {}^{+2.1}_{-0.3} \pm 1.1$ 
&  $30.7 \pm 0.7 {}^{+0.9}_{-1.1}$
&  $87.2 {}^{+2.8}_{-1.0} {}^{+2.0}_{-2.2}$
\\ 
14 TeV &  $154 {}^{+4}_{-1} \pm 3$ 
&  $94 {}^{+2}_{-1} {}^{+2}_{-3}$ 
& $248 {}^{+6}_{-2} {}^{+5}_{-6}$
\end{tabular}}
\caption{NNLO approximate $t$-channel single-top and single-antitop cross sections with $m_t=173$ GeV. The first uncertainty is from scale variation between 
$m_t/2$ and $2m_t$ and the second uncertainty is from the MSTW2008 NNLO pdf \cite{MSTW} at 90\% CL.}
\label{tchannel}
\end{table}

Table \ref{tchannel} lists the $t$-channel single-top and single-antitop cross sections, and their sum,  at 7, 8, and 14 TeV LHC energies, for a top quark mass  $m_t=173$ GeV. The central results are with $\mu_F=\mu_R=m_t$ and the first uncertainty is due to scale variation over the interval $m_t/2$ to $2m_t$, while the second uncertainty denotes the pdf errors using MSTW2008 NNLO pdf at 90\% CL.   
The theoretical ratio $\sigma(t)/\sigma({\bar t})= 1.88{}^{+0.11}_{-0.09}$ at 
7 TeV compares well with the ATLAS result of $1.81{}^{+0.23}_{-0.22}$ \cite{ATLAStchratio}. 

In addition to the total cross section, the top-quark $p_T$ distribution 
in $t$-channel production is of interest and has been calculated at NLO 
in \cite{BWHL,CFMT,SYMC,FGMS,FRT}. More recently, approximate NNLO results 
based on NNLL resummation appeared in \cite{NKtchpt} 
(for another approach based on SCET, see \cite{WLZ}).
The right plot of Fig. \ref{tchtop} shows the theoretical results for $t$-channel top-quark $p_T$ distributions at LHC energies \cite{NKtchpt}.

\subsection{$s$-channel production}

\begin{figure}
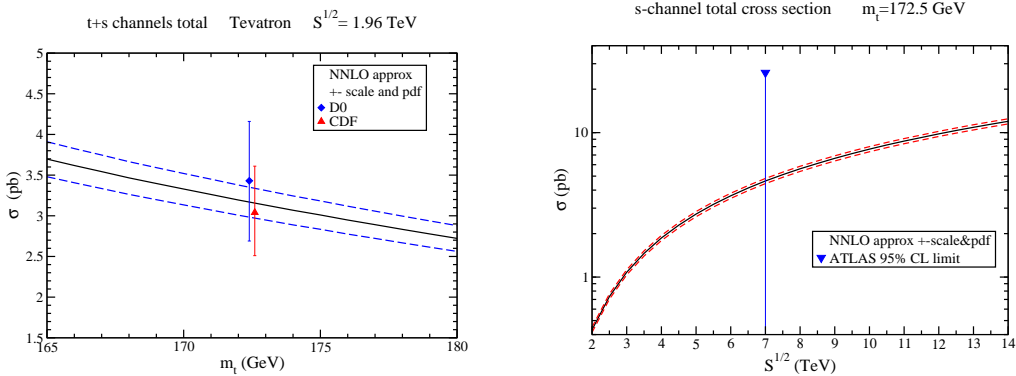

\centerline{\includegraphics[width=0.42\textwidth]{tchschtevplot.eps}
\hspace{10mm}\includegraphics[width=0.42\textwidth]{schtotalSlhcplot.eps}}
\caption{(Left) $t$ and $s$ channel combined cross sections compared with Tevatron data. (Right) $s$-channel cross section at LHC energies.}
\label{tchsch}
\end{figure}

\begin{table}
\centerline{\begin{tabular}{c|c|c|c}
LHC  & $t$ &  
${\bar t}$ &  Total (pb) \\ 
\hline
 7 TeV  &  $3.14 \pm 0.06 {}^{+0.12}_{-0.10}$
& $1.42 \pm 0.01 {}^{+0.06}_{-0.07}$
& $4.56 \pm 0.07 {}^{+0.18}_{-0.17}$
\\
8 TeV  & $3.79 \pm 0.07 \pm 0.13$
& $1.76 \pm 0.01 \pm 0.08$
& $5.55 \pm 0.08 \pm 0.21$
\\ 
14 TeV &  $7.87 \pm 0.14 {}^{+0.31}_{-0.28}$ 
& $3.99 \pm 0.05 {}^{+0.14}_{-0.21}$
& $11.86 \pm 0.19 {}^{+0.45}_{-0.49}$
\end{tabular}}
\caption{NNLO approximate $s$-channel single-top and single-antitop cross sections with $m_t=173$ GeV. The first uncertainty is from scale variation between $m_t/2$ and $2m_t$ and the second uncertainty is from the MSTW2008 NNLO pdf \cite{MSTW} at 90\% CL.}
\label{schannel}
\end{table}

We continue with single top quark production in the $s$-channel.  
The NLO corrections were calculated in \cite{BWHL}. 

The soft anomalous dimension matrix for this process has been calculated at one and two loops \cite{NKsingletopTev,NKsch}. The first element of this $2 \times 2$ matrix for $s$-channel single top production at one loop is 
\cite{NKsingletopTev,NKsch}
\beqa
\Gamma_{S\, s-11}^{(1)}=C_F \left[\ln\left(\frac{s-m_t^2}{m_t\sqrt{s}}\right)
-\frac{1}{2}\right]
\nonumber
\eeqa
and at two loops it is \cite{NKsch}
\beqa
\Gamma_{S\, s-11}^{(2)}=\frac{K}{2} \Gamma_{S\, s-11}^{(1)}
+C_F C_A \frac{(1-\zeta_3)}{4} \, .
\nonumber 
\eeqa

Table \ref{schannel} shows the single top and antitop $s$-channel cross sections at the LHC for $m_t=173$ GeV.
The NNLO approximate corrections provide an enhancement over NLO (with the same pdf) of $\sim 10$\%.  

In the left plot of Fig. \ref{tchsch} the sum of the $t$ and $s$-channel cross
sections at the Tevatron are displayed and compared with D0 \cite{D0tands} 
and CDF \cite{CDFtands} data; the agreement is very good.  The right plot of Fig. \ref{tchsch} shows the $s$-channel cross section as a function of LHC energy together with the current limit from ATLAS \cite{ATLASsch}.

\begin{table}
\centerline{\begin{tabular}{c|c}
Tevatron & Total (pb) at 1.96 TeV \\ 
\hline
$t$-channel  & $2.08 {}^{+0.00}_{-0.04} \pm 0.12$ 
\\
$s$-channel  & $1.05 {}^{+0.00}_{-0.01} \pm 0.06$
\\ 
$t+s$ sum & $3.13 {}^{+0.00}_{-0.05} \pm 0.18$ 
\end{tabular}}
\caption{NNLO approximate $t$ and $s$ channel total cross sections at the Tevatron with $m_t=173$ GeV.}
\label{tsTeV}
\end{table}

Table \ref{tsTeV} shows the single top and antitop $t$-channel and $s$-channel NNLO approximate cross sections at the Tevatron for $m_t=173$ GeV.

\subsection{$tW^-$ production}

\begin{figure}
\centerline{\includegraphics[width=0.35\textwidth]{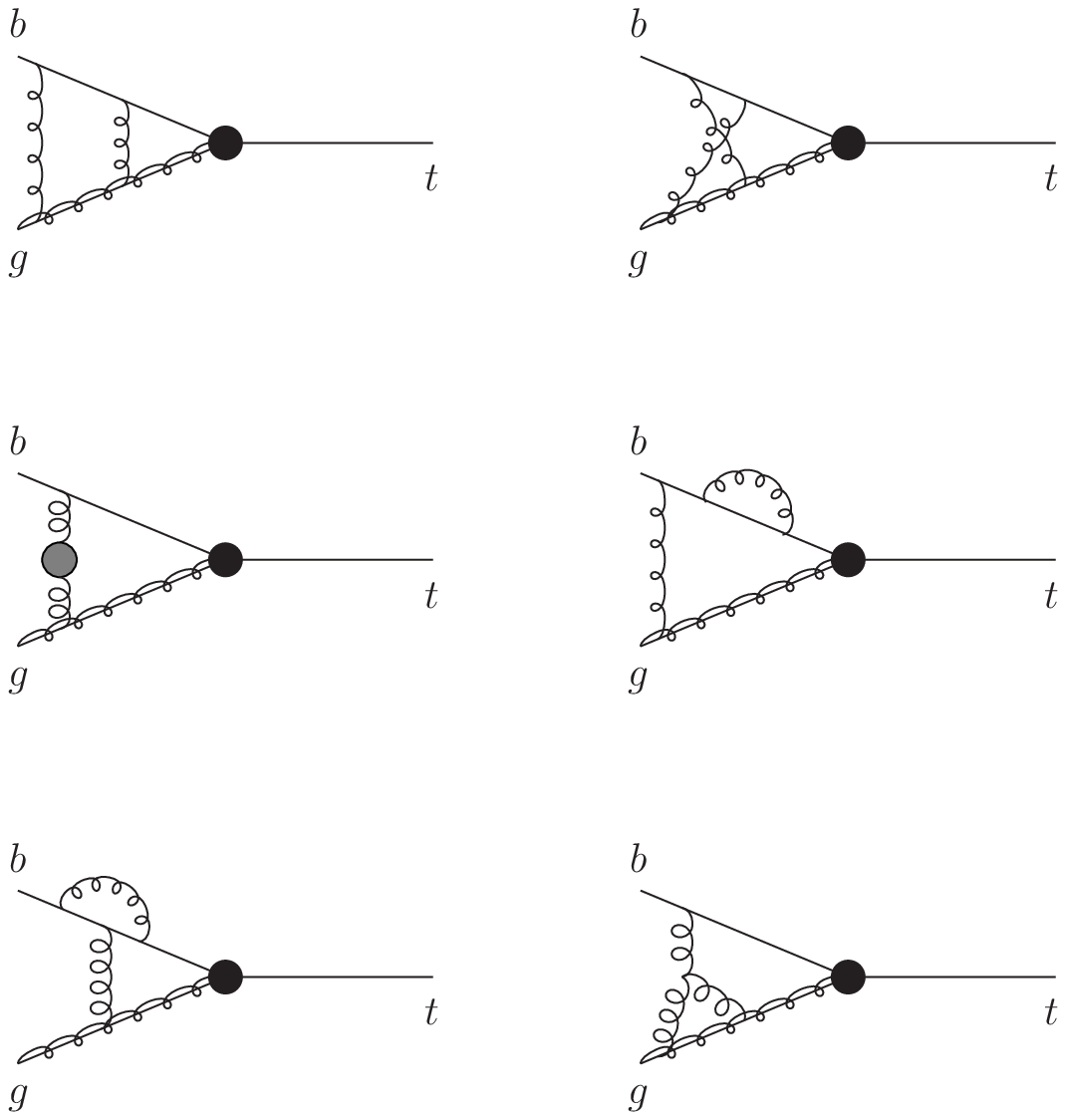}
\hspace{5mm}\includegraphics[width=0.35\textwidth]{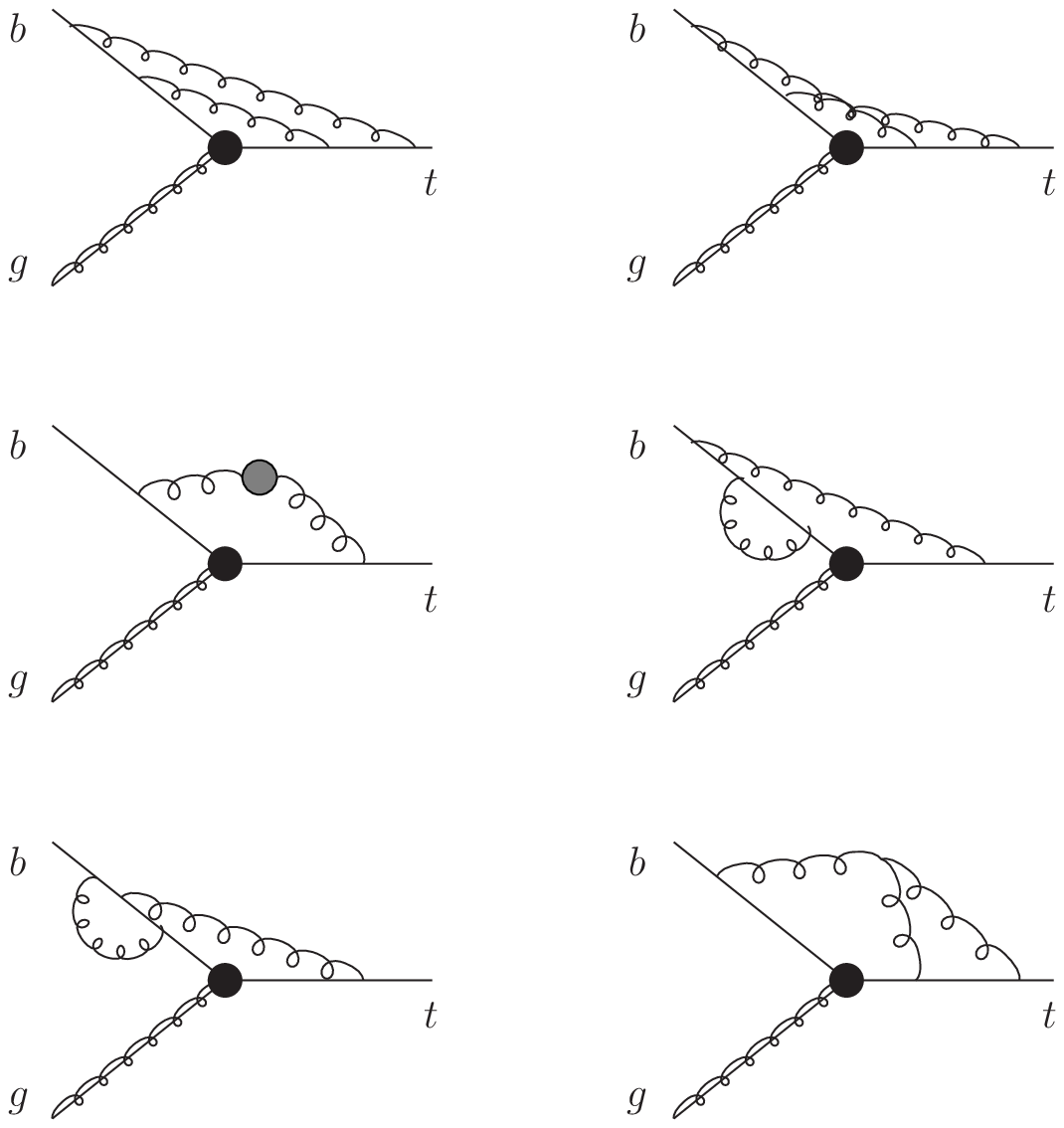}
\hspace{5mm}\includegraphics[width=0.35\textwidth]{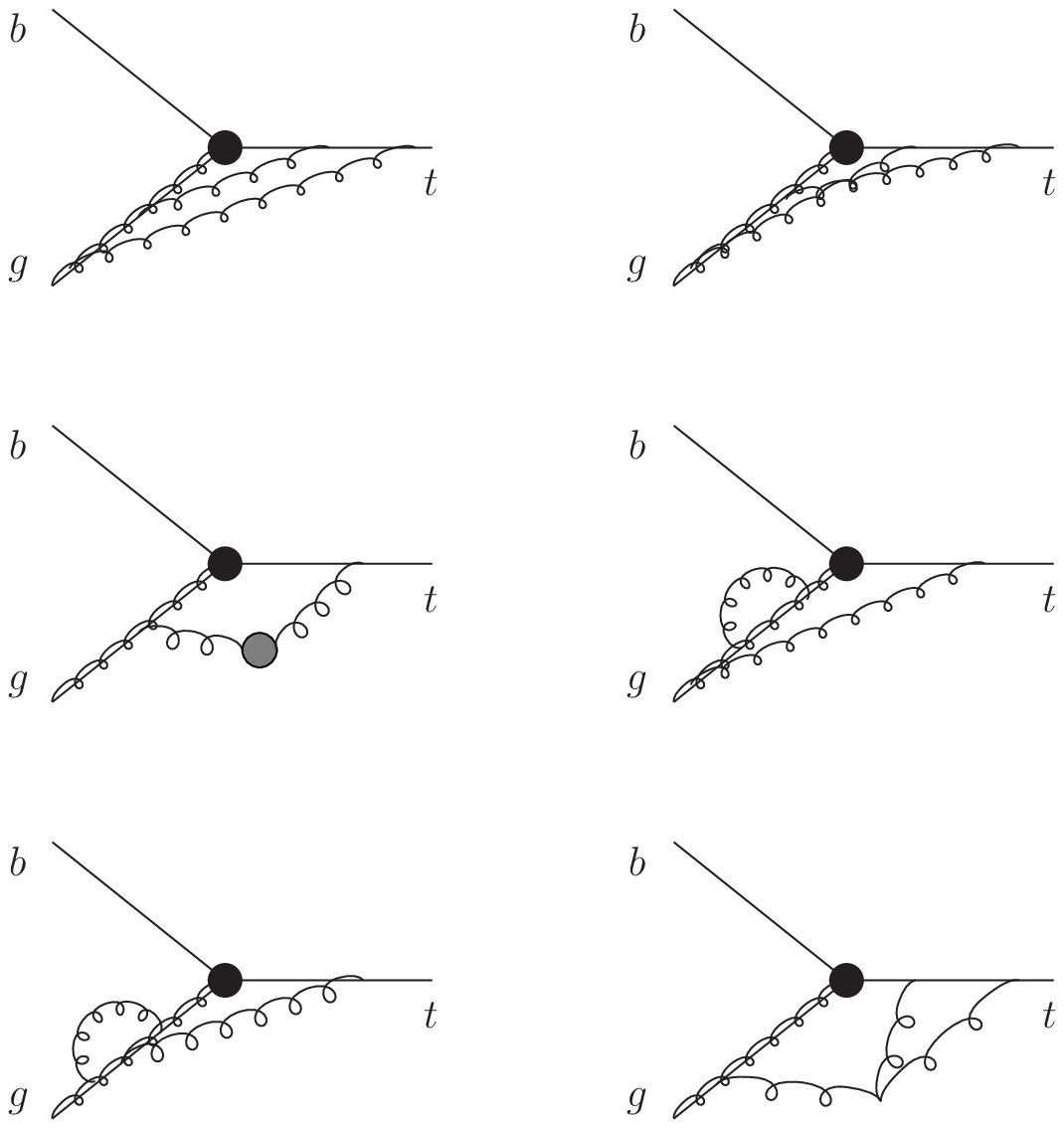}}
\caption{Two-loop eikonal diagrams for $tW$ production.}
\label{tW2}
\end{figure}

We continue with the associated production of a top quark with a $W^-$.
The NLO corrections for this process were calculated in \cite{Zhu}. 
The two-loop eikonal diagrams that contribute to the soft anomalous dimension
are shown in Fig. \ref{tW2} (additional top-quark self-energy graphs also contribute).

The soft anomalous dimension for $bg \rightarrow tW^-$ is given at one loop by 
\cite{NKsingletopTev,NKtW}
\beqa
\Gamma_{S\, tW^-}^{(1)}=C_F \left[\ln\left(\frac{m_t^2-t}{m_t\sqrt{s}}\right)
-\frac{1}{2}\right] +\frac{C_A}{2} \ln\left(\frac{m_t^2-u}{m_t^2-t}\right)
\nonumber
\eeqa
and at two loops by \cite{NKtW}
\beqa
\Gamma_{S\, tW^-}^{(2)}=\frac{K}{2} \Gamma_{S\, tW^-}^{(1)}
+C_F C_A \frac{(1-\zeta_3)}{4} \, .
\nonumber 
\eeqa

\begin{figure}
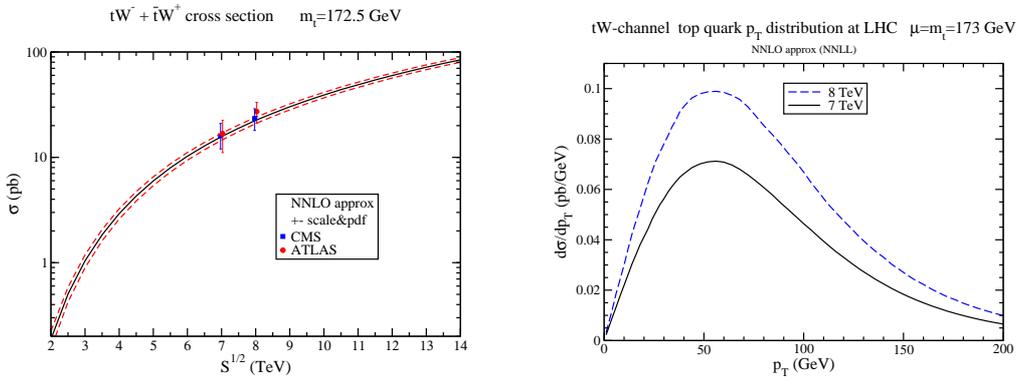

\centerline{\includegraphics[width=0.42\textwidth]{tWtotalSlhcplot.eps}
\hspace{10mm}\includegraphics[width=0.42\textwidth]{pttWlhcplot.eps}}
\caption{Total cross section for $tW$ production (left); top-quark $p_T$ distribution in $tW^-$ production (right).}
\label{pttW}
\end{figure}

The left plot in Fig. \ref{pttW} shows the total $tW$ cross section as a 
function of LHC energy together with LHC data at 7 TeV \cite{ATLAStW7,CMStW7} 
and 8 TeV \cite{ATLAStW8,CMStW8} energy. The agreement of data with theory is 
very good. The right plot displays the top-quark $p_T$ distribution in $tW^-$ 
production at LHC energies.

\begin{table}
\centerline{\begin{tabular}{c|c}
LHC & $tW^-$ (pb) \\ 
\hline
7 TeV  & $7.8 \pm 0.2 {}^{+0.5}_{-0.6}$ 
\\
8 TeV  & $11.1 \pm 0.3 \pm 0.7$ 
\\ 
14 TeV & $41.8 \pm 1.0 {}^{+1.5}_{-2.4}$ 
\end{tabular}}
\caption{NNLO approximate $tW^-$ production cross sections with $m_t=173$ GeV.}
\label{tWchannel}
\end{table}

Table \ref{tWchannel} shows the cross sections for $tW^-$ production at LHC energies for a top quark mass $m_t=173$ GeV.
The NNLO approximate corrections increase the NLO cross section by $\sim 8$\%.
The cross section for ${\bar t}W^+$ production is identical to that for $tW^-$.

\subsection{Associated production of a top quark with a charged Higgs}

Finally, we consider the production of a top quark in association with a charged Higgs boson \cite{NKtW}. Charged Higgs bosons appear in the Minimal Supersymmetric Standard Model (MSSM) and other two-Higgs doublet models.
The soft anomalous dimension for this process is the same as for $tW$ production.

\begin{figure}
\centerline{\includegraphics[width=0.5\textwidth]{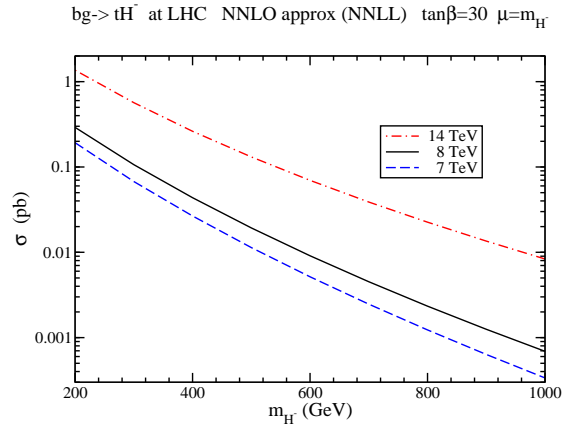}}
\caption{Total cross sections for charged Higgs production in association with a top quark.}
\label{chiggslhcplot}
\end{figure}

Figure \ref{chiggslhcplot} shows the cross section for $tH^-$ production in the MSSM at LHC energies as a function of charged Higgs mass. 
The NNLO approximate corrections increase the NLO cross section by $\sim 15$ to $\sim 20$\%, depending on the charged Higgs mass.

\section{Summary}

In these lectures I have presented higher-order calculations for top-quark 
production in hadronic collisions. I have discussed the resummation of 
soft-gluon corrections for top quark production via different partonic 
channels. NNLL resummation is achieved via two-loop eikonal calculations 
of soft anomalous dimension matrices.
NNLO approximate results for the $t {\bar t}$ production cross section 
and the top quark $p_T$ and rapidity distributions are in excellent agreement 
with data from the LHC and the Tevatron.
Single top cross sections and $p_T$ distributions have been presented in all 
partonic channels and are also in excellent agreement with collider data. 
The NNLO approximate corrections are very significant and they reduce the 
theoretical errors.

\section*{Acknowledgements}

This material is based upon work supported by the National Science Foundation 
under Grant No. PHY 1212472.

% ****************************************************************************
% BIBLIOGRAPHY AREA
% ****************************************************************************

\begin{footnotesize}
% IF YOU DO NOT USE BIBTEX, USE THE FOLLOWING SAMPLE SCHEME FOR THE REFERENCES
% ----------------------------------------------------------------------------

% ----------------------------------------------------------------------------

\end{footnotesize}

% ****************************************************************************
% END OF BIBLIOGRAPHY AREA
% ****************************************************************************

\end{document}